\documentclass[a4paper,11pt]{article}	
\pdfoutput=1
\usepackage{dcolumn}

\usepackage{bm}
\usepackage[T1]{fontenc}
\usepackage{graphicx}
\usepackage{amssymb,amsmath}
\usepackage{booktabs}
\usepackage{multirow}
\usepackage{cite,color,url}
\usepackage[dvipsnames]{xcolor}
\def\linkcolor{cyan!70!black}

\usepackage[
colorlinks=true
,urlcolor=\linkcolor
,anchorcolor=\linkcolor
,citecolor=\linkcolor
,filecolor=\linkcolor
,linkcolor=\linkcolor
,menucolor=\linkcolor
,linktocpage=true
,pdfproducer=medialab
,pdfa=true
]{hyperref}

\usepackage{slashed}
\usepackage{epsfig,psfrag,rotating,soul}
\usepackage{rotfloat}
\usepackage[font={small}]{caption}
\usepackage{colortbl}
\usepackage{soul}
\usepackage{xtab}
\usepackage{tikz}
\usepackage{makecell}

\usepackage{tabularx,threeparttable}

\evensidemargin \oddsidemargin
\usepackage{adjustbox}

\allowdisplaybreaks

\let\OLDthebibliography\thebibliography
\renewcommand\thebibliography[1]{
  \OLDthebibliography{#1}
  \setlength{\parskip}{0pt}
  \setlength{\itemsep}{0pt plus 0.3ex}
}

\usepackage{fontawesome5}
\makeatletter
\newcommand{\github}[1]{%
   \href{#1}{\faGithubSquare}%
}
\makeatother

\usepackage{lipsum} 

\usepackage{tikz,xcolor,hyperref}

\definecolor{lime}{HTML}{A6CE39}
\DeclareRobustCommand{\orcidicon}{%
	\begin{tikzpicture}
	\draw[lime, fill=lime] (0,0) 
	circle [radius=0.16] 
	node[white] {{\fontfamily{qag}\selectfont \tiny ID}};
	\draw[white, fill=white] (-0.0625,0.095) 
	circle [radius=0.007];
	\end{tikzpicture}
	\hspace{-3mm}
}

\foreach \x in {A, ..., Z}{%
	\expandafter\xdef\csname orcid\x\endcsname{\noexpand\href{https://orcid.org/\csname orcidauthor\x\endcsname}{\noexpand\orcidicon}}
}

\begin{document}

\begin{titlepage}

\thispagestyle{empty}

\def\thefootnote{\fnsymbol{footnote}}

\begin{flushright}
IFT-UAM/CSIC-26-94
\end{flushright}

\vspace*{0.5cm}

{\Large\flushleft\sffamily\bfseries\par
Two-Stage Machine Learning Strategy for Scalar and Vector \\ [0.25em] Leptoquark Discrimination at the LHC
}

\vspace{0.5cm}
\hrule height 1.5pt
\vspace{0.5cm}

{\bfseries\raggedright\sffamily\par
Ernesto~Arganda\orcidA{}$^{1}$%
\footnote{{\tt \href{mailto:ernesto.arganda@uam.es}{ernesto.arganda@uam.es}}}%
, Mart\'in~de~los~Rios\orcidB{}$^{2}$%
\footnote{\tt \href{mailto:mdelosrios@unc.edu.ar}{mdelosrios@unc.edu.ar}}%
, Andres D. Perez\orcidC{}$^{3}$%
\footnote{\tt \href{mailto:andres.perez@ib.edu.ar}{andres.perez@ib.edu.ar}}%
, \\ Rosa~M.~Sand\'a Seoane\orcidD{}$^{1}$%
\footnote{{\tt \href{mailto:rosa.sanda@uam.es}{rosa.sanda@uam.es}}}%
\,and Alejandro Szynkman\orcidE{}$^{4}$%
\footnote{{\tt \href{mailto:szynkman@fisica.unlp.edu.ar}{szynkman@fisica.unlp.edu.ar}}}%
}
{\sl\footnotesize
\begin{flushleft}
$^1$Departamento de Física Teórica and Instituto de Física Teórica UAM-CSIC, Universidad Autónoma de Madrid, \\ \, Cantoblanco, 28049 Madrid, Spain

$^2$Instituto de Astronom\'ia Te\'orica y Experimental, CONICET - UNC, Laprida 854, X5000BGR, C\'ordoba, Argentina

$^3$Centro Atómico Bariloche, Instituto Balseiro and CONICET, Av. Bustillo 9500, 8400, S.C. de Bariloche, Argentina

$^4$IFLP, CONICET - Dpto. de F\'{\i}sica, Universidad Nacional de La Plata, C.C. 67, 1900 La Plata, Argentina
\end{flushleft}
}

\vspace{0.5cm}

\renewcommand*{\thefootnote}{\arabic{footnote}}
\setcounter{footnote}{0}

\noindent 
{\sc Abstract:} 
We present a machine learning framework for the characterization of leptoquark (LQ) signals at the Large Hadron Collider, focusing on the discrimination between scalar (SLQ) and vector (VLQ) hypotheses. The method is based on a two-stage inference pipeline that combines a classifier trained to separate Standard Model backgrounds from a mixed LQ signal with a second classifier designed to distinguish between SLQ and VLQ scenarios, using the signal yield inferred from the first-stage classifier to guide the corresponding scalar and vector mass hypotheses. A test statistic is constructed from the classifier outputs and interpreted using reference probability density functions. The approach is applied to realistic LHC final states with hadronically decaying tau leptons, multiple jets, and missing transverse momentum, and its performance is assessed using simulated pseudo-experiments. We show that the proposed strategy provides a robust and statistically consistent procedure to discriminate between SLQ and VLQ signals across a wide range of masses and signal stregths. We find that the spin identification power closely follows the discovery potential, demonstrating that determining the spin nature of a newly discovered LQ does not require substantially larger datasets than those needed for discovery itself.

\end{titlepage}

\vspace{0.5cm}
\hrule height 0.25pt
\vspace{0.5cm}

\tableofcontents

\vspace{1.0cm}
\hrule height 0.25pt
\vspace{0.5cm}

\section{Introduction}
\label{intro}

Exploring phenomena beyond the Standard Model (SM) constitutes a primary objective of the Large Hadron Collider (LHC) programme. Among the many well-motivated extensions of the SM, leptoquarks (LQs)~\cite{Pati:1974yy,Georgi:1974sy,Georgi:1974yf,Fritzsch:1974nn,Pati:1975md,Dimopoulos:1979es,Dimopoulos:1979sp,Eichten:1979ah,Schrempp:1984nj,Wudka:1985ef,Buchmuller:1986iq,Buchmuller:1986zs,Angelopoulos:1986uq,Barbier:2004ez} occupy a particularly compelling position, as they naturally arise in a variety of theoretical frameworks aiming to unify quarks and leptons, such as grand unified theories~\cite{Pati:1974yy,Georgi:1974sy}, composite models~\cite{Pati:1975md,Schrempp:1984nj,Gripaios:2009dq}, and scenarios addressing flavor anomalies~\cite{Hiller:2014yaa,Bauer:2015knc,Dorsner:2016wpm,Capdevila:2017bsm,Buttazzo:2017ixm}. LQs couple directly to quark–lepton pairs, leading to distinctive experimental signatures that can be probed at high-energy hadron colliders~\cite{Queiroz:2014pra,Diaz:2017lit,Bhaskar:2021gsy,Bhaskar:2024wic,Arhrib:2026coy}. Despite extensive LHC searches performed by ATLAS~\cite{ATLAS:2011atv,ATLAS:2011zhi,ATLAS:2012aq,ATLAS:2013oea,ATLAS:2015hsi,ATLAS:2016wab,ATLAS:2019ebv,ATLAS:2019qpq,ATLAS:2020dsf,ATLAS:2020dsk,ATLAS:2020xov,ATLAS:2021oiz,ATLAS:2021yij,ATLAS:2021mla,ATLAS:2021jyv,ATLAS:2022wcu,ATLAS:2023uox,ATLAS:2023kek,ATLAS:2023vxj,ATLAS:2023prb,ATLAS:2024huc} and CMS~\cite{CMS:2010ssz,CMS:2010chx,CMS:2011zfm,CMS:2012iln,CMS:2012bfi,CMS:2012cyn,CMS:2014wpz,CMS:2015nep,CMS:2015xzc,CMS:2015gua,CMS:2016fxb,CMS:2017xcw,CMS:2018svy,CMS:2018qqq,CMS:2018txo,CMS:2018lab,CMS:2018oaj,CMS:2018iye,CMS:2018ncu,CMS:2018yiq,CMS:2020wzx,CMS:2021far,CMS:2022nty,CMS:2022goy,CMS:2022ncp,CMS:2023bdh,CMS:2023qdw,CMS:2024bnj} Collaborations, no conclusive evidence for LQs has been found so far. However, the anticipated increase in integrated luminosity, together with improved analysis techniques, is expected to significantly enhance the sensitivity of these searches~\cite{CMS:2018yke,ATLAS:2025tja}, further motivating the identification and characterization of potential LQ signals.

In recent years, machine learning (ML) techniques have become an essential component of the LHC data analysis toolkit~\cite{Feickert:2021ajf}, enabling significant advances in classification, regression, and anomaly detection tasks. In searches for new physics, ML methods have been widely used to enhance signal--background discrimination, improve the reconstruction of kinematic properties, and exploit high-dimensional correlations beyond the reach of traditional approaches (see, e.g.,~\cite{Arganda:2021azw,Karagiorgi:2021ngt,Arganda:2024tqo,Arganda:2025fhx}). In the context of LQ searches, ML-based classifiers have demonstrated improved sensitivity compared to cut-based or low-dimensional analyses, particularly in complex final states~\cite{Pogwizd:2007sp,Florez:2023jdb,Arganda:2023qni,Ahmed:2024iqx,Qureshi:2024naw,Sahoo:2025kdj,Ghosh:2025gue}. Building on these developments, our previous work~\cite{Arganda:2023qni} introduced a dedicated ML framework designed to distinguish between LQ-induced signals and SM backgrounds.

A related and increasingly important question in the context of new physics searches is the discrimination between competing signal hypotheses once an excess has been identified. Several approaches have been developed to address this problem, ranging from the use of kinematic observables sensitive to spin and angular correlations~\cite{Gao:2010qx,Bolognesi:2012mm} to multivariate and likelihood-based methods grounded in statistical inference~\cite{Neyman:1933wgr,Kondo:1988yd}. More recently, ML techniques have been employed to perform hypothesis testing directly, enabling the construction of powerful test statistics that exploit high-dimensional information from collider events~\cite{Cranmer:2015bka,Brehmer:2018hga,Guest:2018yhq}.  These developments also open the possibility of exploiting high-dimensional information in realistic hadron-collider environments, motivating the design of dedicated inference strategies that account for detector effects and complex final states.

\begin{figure}
  \centering
  \includegraphics[width=0.98\textwidth]{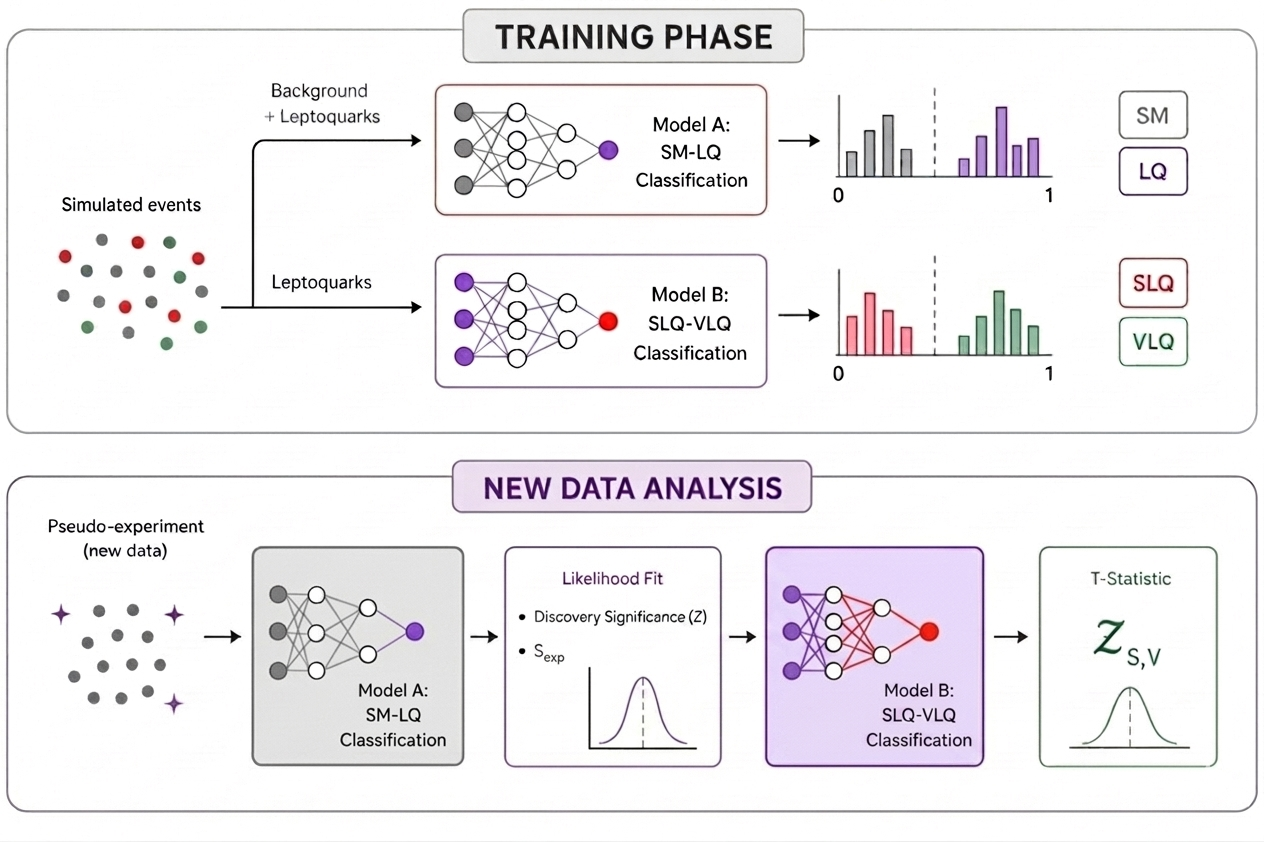}
\caption{Schematic overview of the two-stage machine-learning framework and statistical inference procedure. The upper panel illustrates the training phase of the two ML classifiers: Model A, developed for new physics discovery via SM-LQ separation, and Model B, optimized for signal characterization through scalar versus vector hypothesis discrimination. The lower panel illustrates the inference workflow, showing the sequential application of the pre-trained models to pseudo-experiments or new data. Model~A is first used to identify a signal-enriched event sample and to infer the signal yield through a likelihood fit. The selected events are then passed to Model~B, whose output is used to construct the test statistic and quantify the compatibility of the observed signal with the scalar and vector hypotheses.
}
\label{SLQ_VLQ_pipeline}
\end{figure}

In the event of a discovery consistent with LQ production at the LHC, determining the properties of the new state, and in particular its spin, will be essential~\cite{Zhang:2019jwp,Bandyopadhyay:2020wfv,Qureshi:2024naw}. Distinguishing scalar leptoquarks (SLQs) from vector leptoquarks (VLQs) is essential for identifying the correct ultraviolet completion and understanding the associated dynamics. This task is challenging, as SLQ and VLQ scenarios can lead to very similar experimental signatures once detector effects are taken into account, particularly in complex hadronic final states, thereby limiting the discriminating power of conventional kinematic observables. In this work, we develop a novel ML-based inference pipeline, schematically illustrated in Fig.~\ref{SLQ_VLQ_pipeline}, to address this problem. The approach relies on two classifiers. Model A is a binary classifier trained to distinguish SM events from a mixture of LQ signals, and builds upon the classifier introduced in Ref.~\cite{Arganda:2023qni}, with an improved training based on an extended set of input features. Its output is first used to define a signal-enriched region by applying an appropriate working point, ensuring that the subsequent analysis is performed on events with a high probability of originating from a leptoquark signal. In addition, the full output distribution of Model~A is exploited in a binned-likelihood (BL) fit~\cite{Cowan:2010js} to infer the expected signal yield, which is subsequently provided as an additional input feature to Model~B, since it carries complementary information on the underlying mass scale. Therefore, the second classifier is trained to discriminate between pure SLQ and VLQ hypotheses, implicitly pairing SLQ samples of a given mass with VLQ samples shifted by $\sim400~\mathrm{GeV}$ through the additional signal yield feature, in order to compare scenarios with similar signal rates. The classifier outputs are combined into a test statistic, which is compared with reference probability density functions to construct a log-likelihood ratio and the corresponding statistical significance. This enables a statistically well-defined discrimination between SLQ and VLQ interpretations, as well as an assessment of compatibility with either or both hypotheses. 

The proposed strategy provides a systematic and statistically robust approach to the spin-discrimination problem, with potential applicability to a broader class of BSM scenarios. The use of two dedicated classifiers, rather than a single multi-class model, is motivated by the fundamentally different objectives of the two inference tasks. The first task consists of establishing the presence of a leptoquark signal against the overwhelming SM background, while the second focuses on distinguishing between scalar and vector hypotheses once a signal-like sample has been identified. A sequential approach allows each classifier to optimize a well-defined objective and prevents the spin-discrimination stage from being dominated by signal--background differences. Furthermore, conditioning the second classifier on both a signal-enriched event sample and the signal yield inferred by Model~A enables a more robust extraction of spin-dependent information, closely reflecting the experimental workflow that would naturally follow a potential discovery.

In the present work, we focus on LQs coupled to third-generation fermions. This choice is primarily motivated by phenomenological and experimental considerations. On the one hand, third-generation LQs give rise to distinctive final states involving tau leptons, heavy-flavour jets, and large missing transverse momentum, which provide a rich kinematic structure for multivariate analyses~\cite{ATLAS:2021jyv,CMS:2022nty}. On the other hand, this scenario offers a well-defined and controlled setup to develop and validate the proposed ML strategy. From a theoretical perspective, enhanced couplings to third-generation fermions are also motivated in models addressing the flavor anomalies observed in semileptonic $B$-meson decays~\cite{Hiller:2014yaa,Bauer:2015knc,Buttazzo:2017ixm,LHCb:2024jll,Belle-II:2025yjp}. While the framework introduced in this paper is expected to be applicable to more general flavor structures, the restriction to third-generation couplings allows us to isolate and study the spin-discrimination problem in a representative and experimentally relevant configuration.

The remainder of this paper is organized as follows. In Section~\ref{Ph-frame} we introduce the LQ models considered and the phenomenological assumptions imposed. Section~\ref{simulation} is devoted to the details of the simulation of LQ signals and SM backgrounds, and the subsequent event selection. Section~\ref{ML-stat} describes the machine learning classifiers and the associated statistical tests used in the analysis. The collider setup and event-level analysis are presented in Section~\ref{Results}. Finally, we summarize our results and conclusions in Section~\ref{Conclu}.

\section{LQ Models and Phenomenological Assumptions}
\label{Ph-frame}

In this section we summarise the theoretical framework adopted to describe scalar and vector leptoquarks, together with the phenomenological assumptions underlying the analysis. A defining property of LQs is their ability to mediate direct interactions between quarks and leptons, allowing transitions between both sectors through renormalizable operators. In this work, we focus exclusively on scalar and vector leptoquarks coupled to third-generation SM fermions while conserving baryon and lepton number.

The restriction to third-generation fermions is well motivated both theoretically and phenomenologically. From a theoretical perspective, flavor models aiming to address the observed hints of lepton-flavor non-universality in semileptonic $B$-meson decays typically predict enhanced LQ couplings to third-generation fermions~\cite{Hiller:2014yaa,Bauer:2015knc,Capdevila:2017bsm,Buttazzo:2017ixm,LHCb:2024jll,Belle-II:2025yjp}. In addition, restricting the flavor structure to a single generation naturally suppresses flavor-changing neutral currents and helps evade stringent low-energy constraints~\cite{Dorsner:2016wpm}. From a collider perspective, third-generation final states involving $\tau$ leptons and heavy quarks lead to characteristic signatures with sizable missing transverse energy and high hadronic activity, which are particularly suitable for multivariate analyses and are actively explored in current LHC searches~\cite{ATLAS:2021jyv,CMS:2022nty}. For these reasons, the third-generation scenario provides a theoretically consistent and experimentally well-motivated framework in which to investigate the spin properties of leptoquark signals.

The parameter space of the LQ scenarios considered in this work is constrained by a wide range of experimental searches and low-energy observables. Direct searches at the LHC provide the most stringent bounds on LQ masses for the third-generation scenario under consideration. In particular, recent analyses by the ATLAS and CMS Collaborations exclude SLQs with masses below approximately $1.2$--$1.9~\mathrm{TeV}$, depending on the assumed branching fractions and decay channels~\cite{ATLAS:2021jyv,CMS:2022nty,CMS:2023bdh}. Similar searches for VLQs typically yield stronger limits due to their larger production cross sections, extending to higher masses depending on the coupling structure~\cite{Baker:2019sli,Dorsner:2016wpm}.

Complementary constraints arise from low-energy flavor observables and precision measurements, which restrict the size and flavor structure of the LQ couplings~\cite{Dorsner:2016wpm}. However, in scenarios where the couplings are dominated by third-generation fermions, many of these indirect bounds can be significantly alleviated. As a result, LQ masses in the TeV range remain phenomenologically viable and within the reach of current and future LHC searches. The analysis presented in this work is therefore performed within this experimentally allowed and theoretically consistent region of parameter space.

For the scalar scenario, we adopt the Buchmüller--Rückl--Wyler (BRW) effective framework~\cite{Buchmuller:1986zs}, restricting the interactions to the up-type third-generation state. The relevant Lagrangian can be written as
\begin{align}
\mathcal{L}_{\rm scalar}
&=
g_{3L}\bar{q}^c_L i\tau_2 (\bm{\tau}\cdot\bm{S}_3)l_L
+ 
\left(
h_{2L}\bar{t}_R l_L
+
h_{2R}\bar{q}_L i\tau_2 \tau_R
\right)R_2
+
\tilde{h}_{2L}\bar{b}_R l_L \tilde{R}_2
+\mathrm{c.c.} \,,
\label{eq:Lscalar}
\end{align}
where $q_L$ and $l_L$ denote the left-handed quark and lepton doublets, whereas $t_R$ and $\tau_R$ are the right-handed top-quark and tau-lepton singlets. $S_3$, $R_2$ and $\tilde{R}_2$ represent the scalar LQ fields in the interaction basis. After rotating to the physical basis, the light up-type scalar leptoquark state, denoted by $LQ_3^S$, is obtained from the appropriate admixture of interaction eigenstates and carries electric charge $+2e/3$.

Restricting the couplings to third-generation fermions implies the following dominant decay channels
\begin{align}
LQ_3^S \rightarrow t\nu_\tau,\ b\tau \,.
\end{align}
Consequently, the free parameters of the scalar scenario are the leptoquark mass, $m(LQ_3^S)$, and the branching ratio into a charged lepton,
\begin{align}
\beta = {\rm BR}(LQ_3^S\rightarrow b\tau) \,.
\end{align}

In addition to the scalar case, we also investigate vector leptoquarks. Following Ref.~\cite{Baker:2019sli,ATLAS:2021jyv}, we consider the singlet vector leptoquark $U_1\sim({\bf 3},{\bf 1},2/3)$, whose interaction Lagrangian is given by
\begin{align}
\mathcal{L}_{U_1}
&=
-\frac{1}{2}U_{\mu\nu}^\dagger U^{\mu\nu}
+m_{U}^{2}U_\mu^\dagger U^\mu
-i g_s (1-k_u)
U_\mu^\dagger T^a U_\nu G^{a\mu\nu}
\nonumber\\
&\quad +
\frac{g_U}{\sqrt{2}}
\left[
c_{ij}\,
\bar{Q}_L^i \gamma^\mu L_L^j
\,U_\mu
+
\mathrm{h.c.}
\right] \,,
\label{eq:vectorLQ}
\end{align}
with
\begin{align}
U_{\mu\nu}=D_\mu U_\nu-D_\nu U_\mu \,,
\end{align}
where $G^{a\mu\nu}$ is the gluon field-strength tensor, $T^a$ are the $SU(3)_C$ generators and $D_\mu = \partial_\mu - i g_s G^a_\mu T^a - i \frac{2}{3} g_Y B_\mu$. Typically, two scenarios for the anomalous coupling parameter are customary in phenomenological analyses
\begin{align}
k_u=1 \qquad & \text{minimal coupling scenario},\\
k_u=0 \qquad & \text{gauge model scenario}.
\end{align}
The parameter $k_u$ controls the strength of the interaction between the VLQ and the gluon field-strength tensor. These two benchmark scenarios span a representative range of phenomenological behaviors, in particular affecting both the production cross section and the kinematic distributions of the VLQs, as discussed in detail in Refs.~\cite{Baker:2019sli,DiLuzio:2018zxy}. The value $k_u=1$ corresponds to the minimal coupling scenario, in which the interaction with the SM gauge bosons is only determined by the covariant derivative. On the other hand, $k_u=0$ defines the gauge model scenario, where the VLQ resembles that of a massive gauge boson with anomalous chromo-magnetic interactions. For concreteness, in this work we assume $k_u=1$, which yields a lower production cross section than $k_u=0$. Being closer to the scalar case, this choice presents a more challenging, and thus more conservative, discrimination task. 

Regarding the flavour structure, all entries of the coupling matrix $c_{ij}$ are set to zero except
\begin{align}
c_{33}\neq 0 \,,
\end{align}
so that the vector leptoquark interacts exclusively with third-generation left-handed quark and lepton doublets. Under these assumptions, the dominant vector-leptoquark decay channels are
\begin{align}
U_1 \rightarrow t\nu_\tau,\ b\tau \,.
\end{align}
Then, the free parameters of the vector scenario are the leptoquark mass, $m(LQ_3^V)$, and the branching ratio into a charged lepton, $\beta = {\rm BR}(LQ_3^V\rightarrow b\tau)$.

LQ pair production at the LHC is driven predominantly by QCD interactions. As a consequence, to a very good approximation, the total production cross section depends only on the LQ mass and on its spin, and is largely independent of the Yukawa-like couplings to fermions. This feature allows for a model-independent interpretation of collider searches in terms of the LQ mass and branching fractions.

In contrast, the decay patterns are fully controlled by the fermionic couplings appearing in the Lagrangians expressed in Eqs.~(\ref{eq:Lscalar}) and~(\ref{eq:vectorLQ}). In the present analysis, these are not treated as free parameters. Instead, their effects are encoded in the single parameter $\beta$, defined as the branching ratio into a charged lepton final state. This approach follows the standard experimental strategy and enables a direct comparison with existing collider limits.

Throughout this work, for the assumed values of parameters, the LQ width ($\Gamma(LQ_3^{S/V})$) is sufficiently small compared to its mass ($m(LQ_3^{S/V})$), such that the narrow-width approximation (NWA) is applicable. In this limit, the production and decay of the resonance factorize, and the total cross section can be written as the product of the on-shell production rate and the corresponding branching fractions (see e.g.~\cite{ParticleDataGroup:2024cfk,Uhlemann:2008pm}). Corrections to the NWA are typically of order $\Gamma(LQ_3^{S/V})/m(LQ_3^{S/V})$, although larger deviations may arise in specific scenarios~\cite{Berdine:2007uv}.

For definiteness, we consider a reference scenario with $\beta=0.5$, corresponding to equal branching fractions into charged- and neutral-lepton final states. This choice maximizes the signal yield in mixed final states of the type $t\nu + b\tau$ considered in this analysis and is commonly adopted in experimental searches, where benchmarks with fixed branching fractions are used for the interpretation of results~\cite{ATLAS:2021jyv,CMS:2022nty}. As discussed below, this benchmark will be reinterpreted for arbitrary values of $\beta$ through an appropriate reweighting procedure.

Summarizing, the parameter space explored in this analysis is characterized by the LQ mass and its spin, together with the branching fraction $\beta$. The flavor structure is fixed by imposing $c_{33}\neq 0$ as the only non-vanishing entry, ensuring that both SLQ and VLQ couple exclusively to third-generation fermions. This parameterization allows for a direct comparison between scalar and vector hypotheses in a controlled and phenomenologically well-motivated setup, isolating the effects of the LQ spin on the kinematic properties of the signal. The benchmark mass ranges and their mapping between scalar and vector hypotheses will be detailed in Section~\ref{simulation}, where the simulation setup is introduced.

\section{Simulation Framework and Event Selection}
\label{simulation}

Monte Carlo samples for both signal and background processes were produced at parton level using {\sc MadGraph5\_aMC@NLO}~\cite{Alwall:2014hca} at leading order in QCD, employing the NNPDF2.3 LO parton distribution functions~\cite{Ball:2012cx}. Following the simulation setup of Ref.~\cite{Arganda:2023qni}, event generation was performed for a proton--proton center-of-mass energy of 13~TeV. The corresponding production cross sections were subsequently rescaled to 13.6~TeV and 14~TeV to obtain projections relevant for LHC Run 3 and the HL-LHC program, respectively. The generated events were subsequently interfaced with {\sc Pythia}~\cite{Sjostrand:2014zea,Sjostrand:2007gs} for the simulation of parton showering and hadronization, while detector effects were modeled with {\sc Delphes}~\cite{deFavereau:2013fsa} using the standard ATLAS detector configuration for $\sqrt{s}=13$ and 13.6 TeV, and a HL-LHC configuration for $\sqrt{s}=14$ TeV.

For the signal hypothesis, we considered pair production of third-generation leptoquarks, examining both scalar and vector representations. The decay topology of interest corresponds to one leptoquark yielding a top quark and a neutrino, whereas the second leptoquark decays into a bottom quark and a tau lepton. The implementation of scalar leptoquark interactions was carried out using the UFO model introduced in~\cite{Mandal:2015lca}. For the vector leptoquark scenario with minimal couplings, we employed the corresponding UFO available in~\cite{Baker:2019sli,DiLuzio:2018zxy,Cornella:2021sby}, which provides the necessary interaction structure and model files for event generation. The full process can be written as
\begin{equation}
pp \rightarrow LQ_3^{S/V}\,LQ_3^{S/V}
\rightarrow t\,\nu + b\,\tau \, .
\end{equation}

For background events we considered the main contributions in the single-tau signal region: $t\Bar{t}$ (with 1 real $\tau_{had}$), fake-$t\Bar{t}$ (with no real $\tau_{had}$), single-top, $W+$jets, 
$t\Bar{t}H$ and $t\Bar{t}V$ (with $V=W+Z$). The $Z+$jets and multiboson backgrounds were not considered as they turned out to be negligible for the selection cuts employed in this work.

\begin{table}[t]
\centering
\begin{tabular}{lcc}
\hline
Object & $p_T^{\rm min}$ [GeV] & Pseudorapidity  \\
\hline
$\tau_h$ & 20 & $|\eta|<2.5$, excluding $1.37<|\eta|<1.52$ \\
$j$      & 20 & $|\eta|<2.8$ \\
$b\text{-jet}$      & 20 & $|\eta|<2.5$ \\
$e$      & 10 & $|\eta|<2.47$ \\
$\mu$    & 10 & $|\eta|<2.7$ \\
\hline
\end{tabular}
\caption{Baseline object-selection requirements used throughout the analysis. Here $\tau_h$ denotes a hadronically decaying tau lepton.}
\label{tab:object_selection}
\end{table}

Following our previous analysis~\cite{Arganda:2023qni}, which was based on the ATLAS search~\cite{ATLAS:2021jyv}, we adopt the same object reconstruction and identification requirements defined by the Collaboration. The corresponding kinematic acceptance criteria are summarized in Table~\ref{tab:object_selection}. 
Signal yields are normalized using the same overall normalization factor employed in Ref.~\cite{Arganda:2023qni}, obtained by validating the simulation chain against the ATLAS scalar-leptoquark benchmark~\cite{ATLAS:2021jyv}. Although this factor was originally derived for the scalar scenario by fitting the detector-level $p_T(\tau)$ spectrum, it is also applied to the vector-leptoquark samples in the present analysis. This choice is supported by the similar detector-level $p_T(\tau)$ spectra observed for scalar and vector leptoquark signals, suggesting that the correction mainly accounts for an overall normalization of the simulation chain rather than for shape-dependent effects. Further details can be found in~\cite{Arganda:2023qni}.

The signal region is defined by requiring exactly one hadronically decaying tau lepton ($\tau_h$) and at least two $b$-tagged jets. In addition, events containing isolated electrons or muons are vetoed, and a minimum missing transverse energy ($E_T^{\rm miss}$) requirement of  280 GeV is imposed. This signal region is analogous to the one adopted in our previous study~\cite{Arganda:2023qni} for scalar leptoquarks. The choice is motivated not only by practical considerations related to the event simulation, since the selection requirements defining the single-$\tau$ multi-bin signal region of~\cite{ATLAS:2021jyv}, for instance, are rather restrictive and lead to a very limited number of Monte Carlo events after selection, but more importantly by the multivariate nature of the present analysis. Indeed, multivariate classifiers generally benefit from retaining a larger fraction of the available kinematic information, whereas excessively tight signal regions may inadvertently remove discriminating features that could otherwise enhance the separation between signal and background. We assume that the same strategy can be consistently extended to the vector leptoquark scenario considered in this work. Event selection is summarized in Table~\ref{tab:preselection}.

\begin{table}[t]
\centering
\begin{tabular}{cc}
\hline
Observable & Requirement \\
\hline
$N_{\tau_h}$ & $=1$ \\
$N_{b\text{-jets}}$ & $\geq 2$ \\
$N_{e,\mu}$ & $=0$ \\
$E_T^{\rm miss}$ & $\geq280~\mathrm{GeV}$ \\
\hline
\end{tabular}
\caption{Summary of the selection cuts used in the analysis.}
\label{tab:preselection}
\end{table}

Signal samples for scalar leptoquarks were generated in the mass range $m(LQ_3^{S}) \in [1000,1900]$ GeV, considering benchmark points separated by 100 GeV and fixing the branching fraction parameter to $\beta=0.5$. For vector leptoquarks, the corresponding mass range was chosen as $m(LQ_3^{V}) \in [1400,2300]$ GeV, using the same mass spacing and value of $\beta$. The offset between the scalar and vector mass intervals is motivated by the larger production cross sections of vector leptoquarks. In particular, the selected ranges allow for a meaningful correspondence between scalar and vector benchmark points yielding comparable signal rates in the final state under consideration. This choice is essential for assessing the ability of the proposed analysis to discriminate between scalar and vector leptoquark hypotheses in the event of a potential signal observation. For each benchmark point, a sufficiently large number of events was generated to retain at least $\sim 100$k signal events after applying the selection criteria described above. Likewise, the background samples were produced with enough statistics to obtain a detector-level dataset containing at least $\sim 1$M events after accounting for the relative weights of the different background processes.

Since the signal samples were generated with $\beta=0.5$, both leptoquark decay modes, namely into a quark and a neutrino or into a quark and a charged lepton, contribute to the simulated event sample. These samples can subsequently be reweighted to different branching fractions in order to derive limits in the $m(LQ_3^{S/V})$ versus BR$(LQ_3^{S/V}\rightarrow b\tau)$ plane. To this end, for each value of $m(LQ_3^{S/V})$, additional samples of $\sim50$k events was generated across the interval $\beta\in[0,1]$. These samples are used to determine the relative signal efficiencies and cross sections after selection, enabling the reweighting of the $\beta=0.5$ benchmark to arbitrary branching fractions.

\section{ML Binary Classifiers and Statistical Tests}
\label{ML-stat}

In this section, we describe the procedure developed to distinguish between scalar and vector leptoquarks. The strategy is based on two machine-learning classifiers trained on simulated events, corresponding to Models A and B in the workflow shown in Fig.~\ref{SLQ_VLQ_pipeline}. The first model is designed to discriminate leptoquark signal events, regardless of their spin hypothesis or masses, from the SM background. This constitutes a standard signal-versus-background classification task. In addition to providing signal--background separation, its output can be used to estimate the expected significance of the observation through a BL test to the model response. Events identified as signal candidates are subsequently processed by the second classifier, denoted as Model B. This classifier is trained to distinguish between the scalar and vector leptoquark hypotheses, thereby probing the spin nature of the underlying particle. The determination of whether the observed signal is more compatible with scalar or vector leptoquarks constitutes the main objective of this work.

\begin{table}[ht]
\centering
\small
\begin{tabular}{lllll}
\hline
\multicolumn{5}{c}{\textbf{Object kinematics}}\\
\hline
$p_T(b_1)$ & $\eta(b_1)$ & $\phi(b_1)$ & $p_T(b_2)$ & $\eta(b_2)$ \\
$\phi(b_2)$ & $p_T(\tau_h)$ & $\eta(\tau_h)$ & $\phi(\tau_h)$ & $p_T(j_1)$ \\
$\eta(j_1)$ & $\phi(j_1)$ & $p_T(j_2)$ & $\eta(j_2)$ & $\phi(j_2)$ \\
\hline
\multicolumn{5}{c}{\textbf{Global event observables}}\\
\hline
$E_T^{\mathrm{miss}}$ & $\phi(E_T^{\mathrm{miss}})$ & $H_T$ & $s_T$ & $N_{\mathrm{jets}}$ \\
$N_{b\mathrm{-jets}}$ & & & & \\
\hline
\multicolumn{5}{c}{\textbf{Angular observables}}\\
\hline
$\Delta R(\tau_h,b_1)$ & $\Delta R(\tau_h,b_2)$ & $\Delta R(b_1,b_2)$ & $\Delta R(j_1,b_1)$ & $\Delta R(j_1,b_2)$ \\
$\Delta R(j_1,\tau_h)$ & $\Delta R(j_2,b_1)$ & $\Delta R(j_2,b_2)$ & $\Delta R(j_2,\tau_h)$ & $\Delta R(j_1,j_2)$ \\
$\Delta\phi(E_T^{\mathrm{miss}},b_1)$ &
$\Delta\phi(E_T^{\mathrm{miss}},b_2)$ &
$\Delta\phi(E_T^{\mathrm{miss}},\tau_h)$ &
$\Delta\phi(E_T^{\mathrm{miss}},j_1)$ &
$\Delta\phi(E_T^{\mathrm{miss}},j_2)$ \\
\hline
\multicolumn{5}{c}{\textbf{Mass observables}}\\
\hline
$m(\tau_h,b_1)$ & $m(\tau_h,b_2)$ & $m(b_1,b_2)$ & $m(j_1,b_1)$ & $m(j_1,b_2)$ \\
$m(j_1,\tau_h)$ & $m(j_2,b_1)$ & $m(j_2,b_2)$ & $m(j_2,\tau_h)$ & $m(j_1,j_2)$ \\
$m_T(E_T^{\mathrm{miss}},b_1)$ &
$m_T(E_T^{\mathrm{miss}},b_2)$ &
$m_T(E_T^{\mathrm{miss}},\tau_h)$ &
$m_T(E_T^{\mathrm{miss}},j_1)$ &
$m_T(E_T^{\mathrm{miss}},j_2)$ \\
$m_T(b_1,b_2)$ & $m_T(j_1,j_2)$ & & & \\

\hline
\end{tabular}
\caption{List of the 53 input features used for training Models A and B. The variables are grouped according to their physical interpretation: object-level kinematics, global event observables, angular observables and mass observables.}
\label{tab:variables}
\end{table}

A comprehensive set of kinematic, angular, and mass-related observables is used as input to both Models A and B. The low-level features include the transverse momentum, pseudorapidity, and azimuthal angle, $(p_T,\eta,\phi)$, of the reconstructed hadronic tau lepton, $\tau_h$, the two leading $b$-tagged jets, $b_1$ and $b_2$, and the two leading non-$b$-tagged jets, $j_1$ and $j_2$. Since the presence of non-$b$-tagged jets is not required by the selection criteria in Table~\ref{tab:preselection}, events with fewer than two such jets can occur. In these cases, the features associated with the missing jets are zero-padded. In practice, however, events with fewer than two non-$b$-tagged jets are rare. Missing transverse momentum information is incorporated through $E_T^{\rm miss}$ and $\phi(E_T^{\rm miss})$. In addition, other global event observables are considered, namely the non-$b$-tagged jet multiplicity $N_{\rm jets}$, the number of $b$-tagged jets $N_{b\rm{-jets}}$, the hadronic activity
\begin{equation}
H_T = \sum_{i=1}^{N_{\rm jets}+N_{b\rm{-jets}}} p_T(J_i) \,,
\end{equation}
where $J_i$ denotes any reconstructed jet (regardless of its $b$-tagging status), and the variable
\begin{equation}
s_T = p_T(\tau_h) + p_T(J_1) + p_T(J_2) \,,
\end{equation}
where $J_1$ and $J_2$ represent the two highest-$p_T$ jets in the event, irrespective of $b$-tagging.
As shown in Ref.~\cite{Arganda:2023qni}, these observables already provide near-optimal discrimination between the signal and the SM backgrounds. 

However, since Model B is additionally required to distinguish between scalar and vector leptoquark hypotheses, a richer set of high-level observables is included. These variables encode complementary information on the event topology, the spatial correlations among the final-state particles, and the characteristic mass scales of the underlying process. In particular, we consider angular separations between pairs of reconstructed objects $X$ and $Y$ through the distance measure
\begin{equation}
\Delta R(X,Y)=\sqrt{\left[\eta(X)-\eta(Y)\right]^2+\left[\phi(X)-\phi(Y)\right]^2} \,,
\end{equation}
as well as azimuthal-angle differences $\Delta\phi$ involving $E_T^{\rm miss}$. We also include invariant masses of particle pairs, defined generically as
\begin{equation}
m(X,Y)=\sqrt{\left(p(X)+p(Y)\right)^2}
=\sqrt{2\,p_T(X)\,p_T(Y)\left[\cosh\!\left(\Delta\eta(X,Y)\right)-\cos\!\left(\Delta\phi(X,Y)\right)\right]} \,,
\end{equation}
where $p(X)$ and $p(Y)$ denote the corresponding four-momenta. Furthermore, transverse-mass observables are employed. For a reconstructed object $X$, the transverse mass with respect to the missing transverse momentum is defined as
\begin{equation}
m_T(X)\equiv
m_T\!\left(p_T(X),E_T^{\rm miss}\right)
=
\left[
2\,p_T(X)\,E_T^{\rm miss}
\left(1-\cos\Delta\phi(X,E_T^{\rm miss})\right)
\right]^{1/2} \,.
\end{equation}

The complete list of the 53 input features used for training both models is summarized in Table~\ref{tab:variables}. Together, these observables provide sensitivity not only to the presence of a leptoquark signal, but also to the spin-dependent kinematic patterns that differentiate scalar from vector leptoquark production and decay.

\subsection{Model A: Leptoquarks vs SM background}
\label{modelA}

For Model A, whose purpose is to discriminate leptoquark-induced events from SM backgrounds, we trained a single supervised event-level classifier using the {\tt XGBoost} framework~\cite{Chen:2016btl,Chen:2016:XST:2939672.2939785}. The dataset consists of $1$M signal events and $1$M background events, resulting in a balanced binary classification problem. The data were divided into three datasets: $55\%$ for training, $20\%$ for validation, and the remaining $25\%$ for testing. The signal class was constructed by combining scalar and vector leptoquark samples with equal representation. Furthermore, all benchmark mass points were included with identical statistical weight, ensuring a uniform coverage of the considered mass ranges (sampled in steps of $100$~GeV). The background class is composed of the processes described in Section~\ref{simulation}. Their relative fractions in the training sample are chosen according to their expected contributions after applying the event-selection requirements summarized in Table~\ref{tab:preselection}.

\begin{figure}
  \centering
  \includegraphics[width=0.5\textwidth]{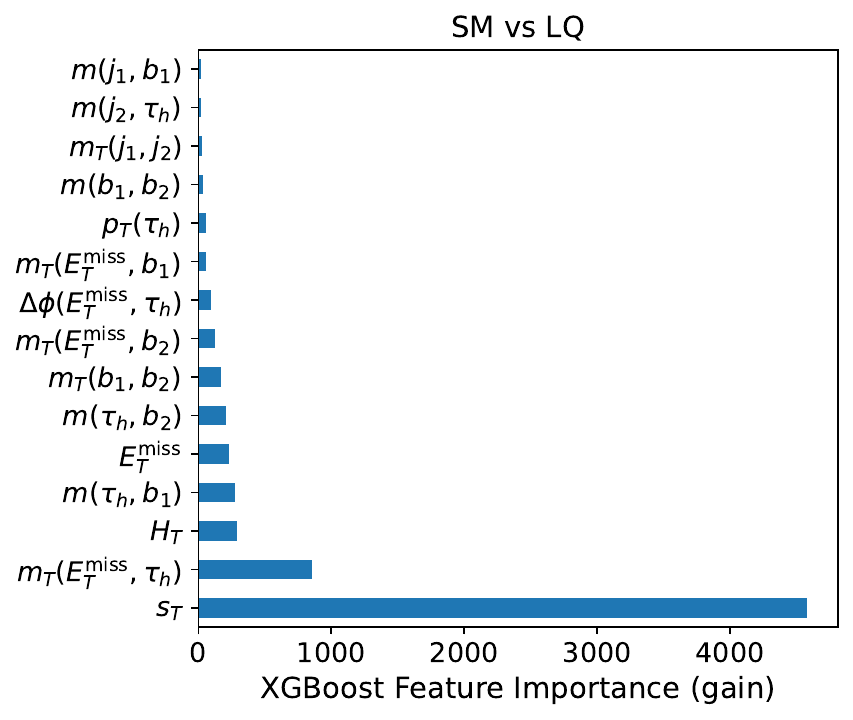}
  \includegraphics[width=0.48\textwidth]{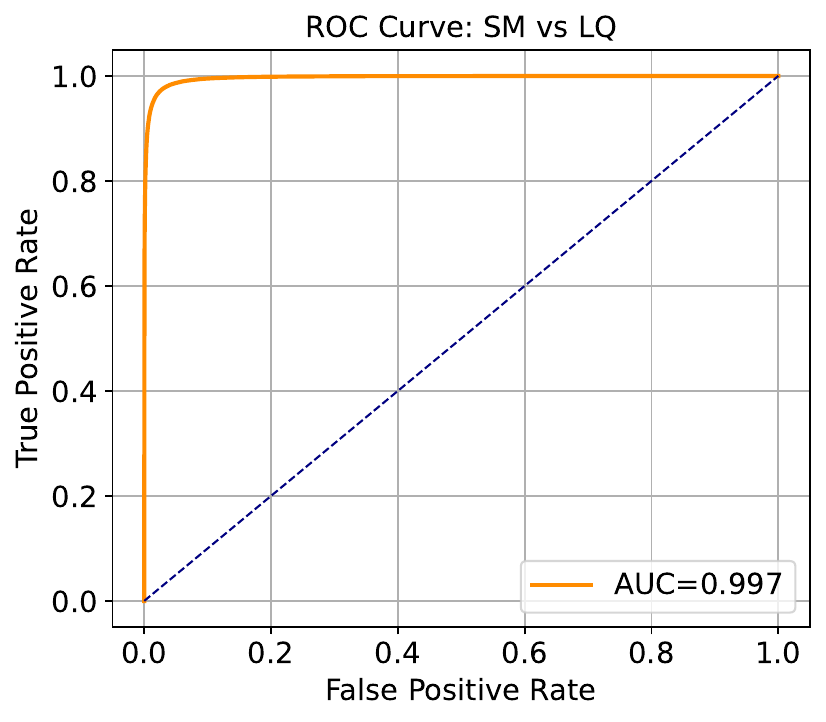}
\caption{
Feature-importance ranking for Model~A, computed using the gain metric of the {\tt XGBoost} classifier (left panel). The score quantifies the relative contribution of each input variable to the LQ--SM discrimination task. ROC curve for Model~A (right panel). The corresponding area under the curve (AUC) of 0.997 demonstrates the excellent separation power achieved between signal and background events.
}
\label{XGBoost-outputs}
\end{figure}

\begin{figure}
  \centering
  \includegraphics[width=0.49\textwidth]{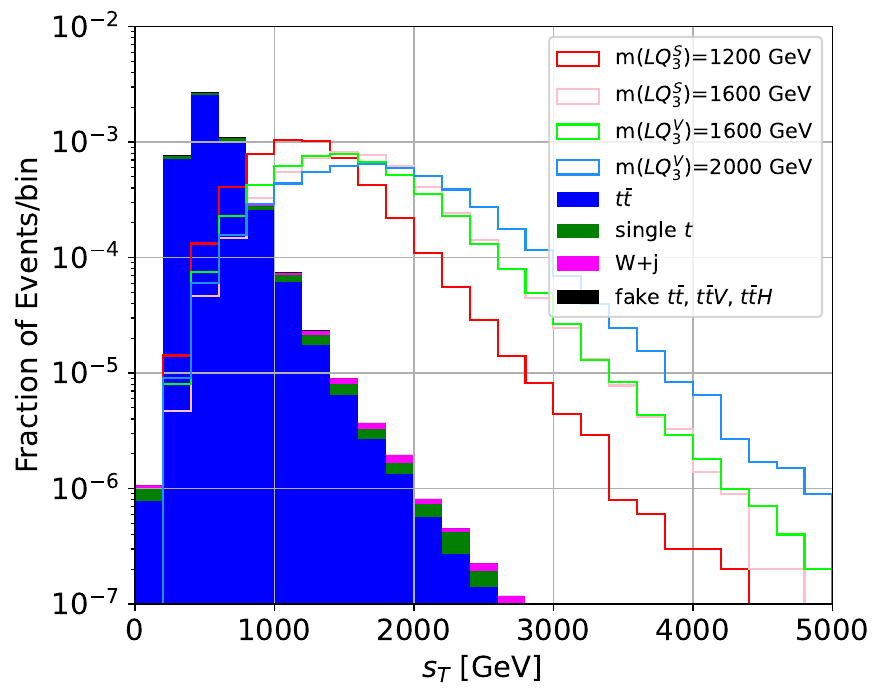}
  \includegraphics[width=0.49\textwidth]{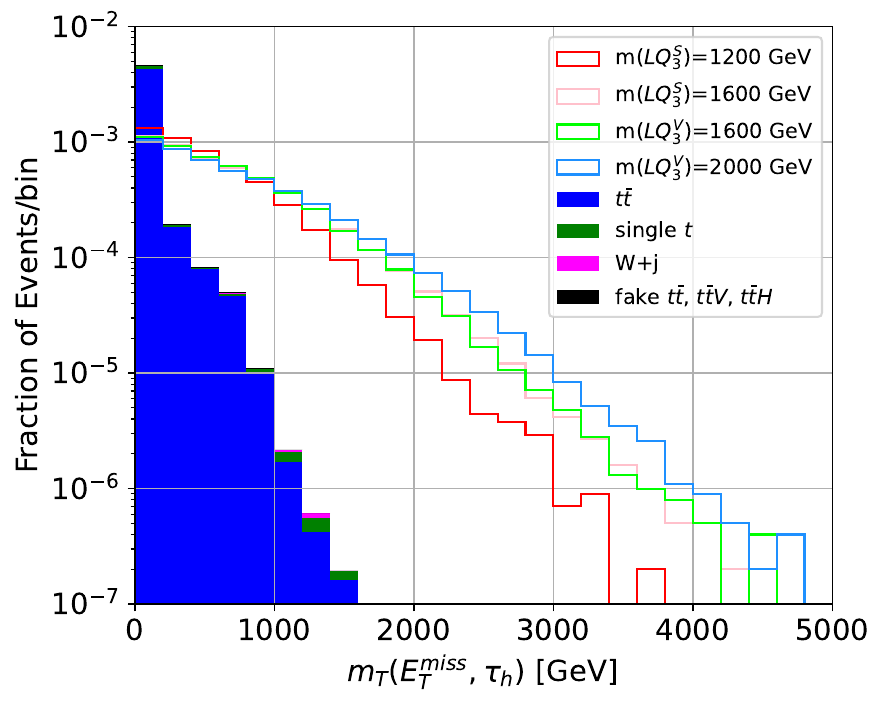}\\
  \includegraphics[width=0.49\textwidth]{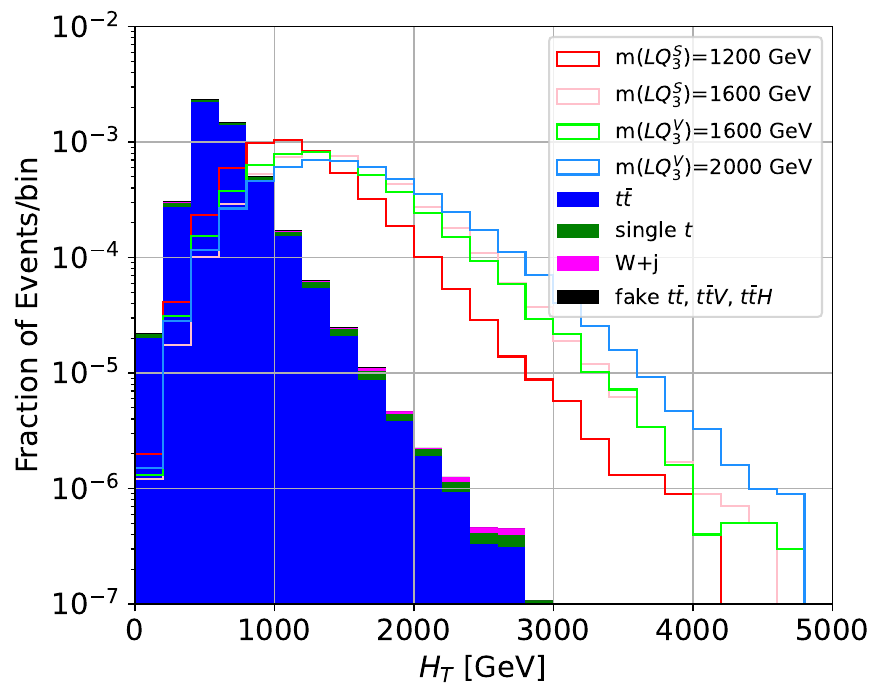}
  \includegraphics[width=0.49\textwidth]{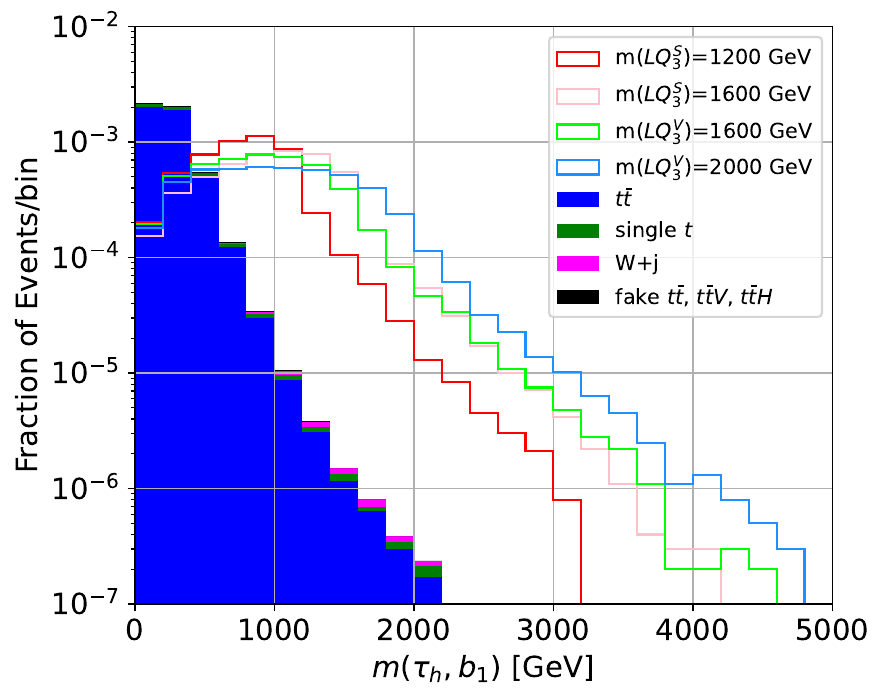}
\caption{Distributions of the four most important input variables for the Model A LQ--SM discrimination: $s_T$ (upper-left panel), $m_T(E_T^{\rm miss},\tau_h)$ (upper-right panel), $H_T$ (lower-left panel), and $m(\tau_h, b_1)$ (lower-right panel). The signal distributions (solid lines) correspond to representative SLQ benchmarks with $m(LQ_3^S)=1200$ and $1600$~GeV and VLQ benchmarks with $m(LQ_3^V)=1600$ and $2000$~GeV, assuming $\beta=0.5$ in all cases. The stacked histograms show the expected SM background contributions after event selection. Minor backgrounds, namely fake-$t\bar{t}$, $t\bar{t}V$, and $t\bar{t}H$, are combined into a single category.}
\label{distributions}
\end{figure}

The feature-importance ranking obtained from the trained classifier is shown in the left panel of Fig.~\ref{XGBoost-outputs}. The importance score is computed using the gain metric, which quantifies the average improvement in the decision trees provided by each feature. Variables with larger gain values contribute more significantly to the classification performance. The most relevant observables are found to be $s_T$, $m_T(E_T^{\rm miss},\tau_h)$, $H_T$, and $m(\tau_h, b_1)$, indicating that the overall energy scale of the event and its missing-momentum structure provide the strongest discrimination between signal and background. The distributions of the most important input variables are presented in Fig.~\ref{distributions}. We observe that signal samples corresponding to different benchmark masses and to both scalar and vector leptoquark hypotheses exhibit remarkably similar behaviors. In contrast, the SM backgrounds display significantly different distributions, explaining the excellent separation power achieved by the classifier. This characteristic is particularly important for the subsequent analysis, since it demonstrates that Model A learns generic characteristics of leptoquark production rather than features associated with a specific mass point or spin hypothesis.

The corresponding receiver operating characteristic (ROC) curve is shown in the right panel of Fig.~\ref{XGBoost-outputs}. The classifier achieves an area under the ROC curve (AUC) of $0.997$, demonstrating an outstanding discrimination capability between signal and background events. Such a high performance indicates that the multidimensional information encoded in the selected observables can be efficiently compressed into a single powerful discriminant. The Model A classifier output, denoted by $o_A(x)\in[0,1]$, quantifies the degree to which an event resembles the signal hypothesis. Values close to $o_A(x)=1$ correspond to highly signal-like events, whereas values near $o_A(x)=0$ are characteristic of SM backgrounds. This mapping effectively transforms the original high-dimensional classification problem into a one-dimensional observable while retaining nearly all of the available discriminating information. 

The excellent performance of Model A is not entirely unexpected. In our previous study~\cite{Arganda:2023qni}, an analogous {\tt XGBoost} architecture, developed for the same $t\nu+b\tau$ final state, achieved an AUC of 0.992 in distinguishing scalar-leptoquark signals from SM backgrounds using only object-level kinematic variables together with a subset of the global event observables employed in this work. In the present analysis, the classifier is provided with a significantly richer set of inputs, including additional high-level observables sensitive to event topology and characteristic mass scales, making a further improvement in performance a natural outcome. More importantly, the results demonstrate that the conclusions reached in Ref.~\cite{Arganda:2023qni} for scalar leptoquarks remain valid when vector leptoquark signals are included. This can be traced back to the remarkably similar kinematic behavior exhibited by scalar and vector leptoquark events, as illustrated in Fig.~\ref{distributions}, indicating that the classifier is primarily exploiting generic features of leptoquark production rather than spin-dependent effects.

The output of Model A is subsequently used to perform a statistical inference of the signal presence. For this purpose, we employ the traditional BL approach. The classifier response is divided into bins, and the expected numbers of signal and background events are computed in each bin. A likelihood function is then constructed as the product of Poisson probabilities associated with the observed event counts. Two complementary statistical tests are then performed within the profile-likelihood framework. First, the discovery significance is evaluated by testing the background-only hypothesis against the signal-plus-background hypothesis, allowing us to determine the integrated luminosity required to achieve evidence ($3\sigma$) or discovery ($5\sigma$). Second, we evaluate the exclusion power ($2\sigma$) by testing the signal-plus-background hypothesis against the background-only hypothesis, using the corresponding profile-likelihood-ratio test statistic. Both tests are implemented following the asymptotic formalism of Ref.~\cite{Cowan:2010js}.

To ensure a statistically robust implementation of the BL procedure, the classifier output is divided into the maximum number of equally sized bins allowed by requiring at least five expected background events per bin. We adopt the same number of bins used in our previous analysis~\cite{Arganda:2023qni}, for which the stability of the results with respect to the binning choice was explicitly verified. Although the larger event yields expected at $\sqrt{s}=14$~TeV would in principle allow for a finer binning, we retain the same number of bins to ensure a consistent comparison with the previous study; this choice is therefore conservative. The sensitivity and corresponding statistical significances are evaluated using 1000 pseudo-experiments. We have verified that increasing the number of pseudo-experiments to 5000 does not lead to any appreciable change in the resulting estimates, indicating that the procedure is numerically stable.

Once the signal significance has been evaluated, the output of Model A is also used to define a signal-enriched region. This is achieved by selecting events with classifier scores above a given threshold called working point. Higher thresholds correspond to increasingly pure signal samples, at the expense of a reduced signal efficiency. Unless otherwise stated, we adopt a reference working point of 0.95, which strongly suppresses the residual SM background while preserving a significant fraction of the signal events. As will be shown and discussed later in Sec.~\ref{sec:threshold_dependence}, the final results exhibit only a mild dependence on the precise choice of this threshold. The events surviving this requirement are subsequently used to evaluate Model B, whose goal is to determine whether the observed signal is more compatible with the scalar or vector leptoquark hypothesis. Since the selected sample is expected to contain a negligible background contamination, this second classification stage effectively focuses on the spin characterization of the potential leptoquark signal.

Last but not least, we think it is important to make a point about the use of binned and unbinned methods. In~\cite{Arganda:2023qni} we demonstrated that using kernel density estimators (KDEs) to fit likelihood functions~\cite{Arganda:2022zbs} improved the sensitivity of scalar LQ masses to around 200–300 GeV compared to the binned method~\cite{Arganda:2022qzy, Arganda:2022mrd}, at the cost of increased computation time. However, in the present work, our Model A with BL functions performs exceptionally well for reliably separating signal from background, with significance values at the discovery level, making the use of the unbinned KDE method counterproductive. We conceive of this first Model A as a kind of signal trigger, which we then process with our Model B to attempt to discern the vector or scalar nature of the LQs.

\subsection{Model B: Scalar vs Vector Leptoquarks}
\label{modelB}

For Model B, whose purpose is to discriminate between scalar and vector third-generation leptoquark hypotheses, we trained a second supervised event-level classifier using the {\tt XGBoost} framework. Unlike Model A, which performs a signal-versus-background classification, Model B is designed to characterize the nature of a potential leptoquark signal by determining whether the observed events are more compatible with a scalar or a vector leptoquark interpretation. 

The dataset consists of approximately $500$k scalar-leptoquark events and $500$k vector-leptoquark events, resulting in a balanced binary-classification problem. The data were divided into three datasets: training, validation, and test, following the same proportions as in Model A. The classifier is trained using the full simulated signal samples after applying the baseline event-selection criteria summarized in Table~\ref{tab:preselection}, without imposing any additional requirement based on the output of Model~A. This choice maximizes the available training statistics and allows Model~B to learn the intrinsic kinematic differences between scalar and vector leptoquark events over the full signal phase space. We verified that applying a moderate signal-enrichment requirement based on the output of Model~A during the training stage of Model~B leads to negligible changes in the classifier performance. This indicates that the signal events with more background-like kinematics carry little weight in the optimization of Model~B and do not significantly affect its ability to discriminate between the two spin hypotheses. The working-point requirement of Model~A is instead applied exclusively during the evaluation stage. In a realistic analysis, Model~B is used only after a signal-like event sample has been selected by Model~A, ensuring that the spin-discrimination procedure operates on the same event sample that would be available following the observation of a potential leptoquark signal. This sequential inference strategy is illustrated in Fig.~\ref{SLQ_VLQ_pipeline}.

\begin{figure}
  \centering
  \includegraphics[width=0.49\textwidth]{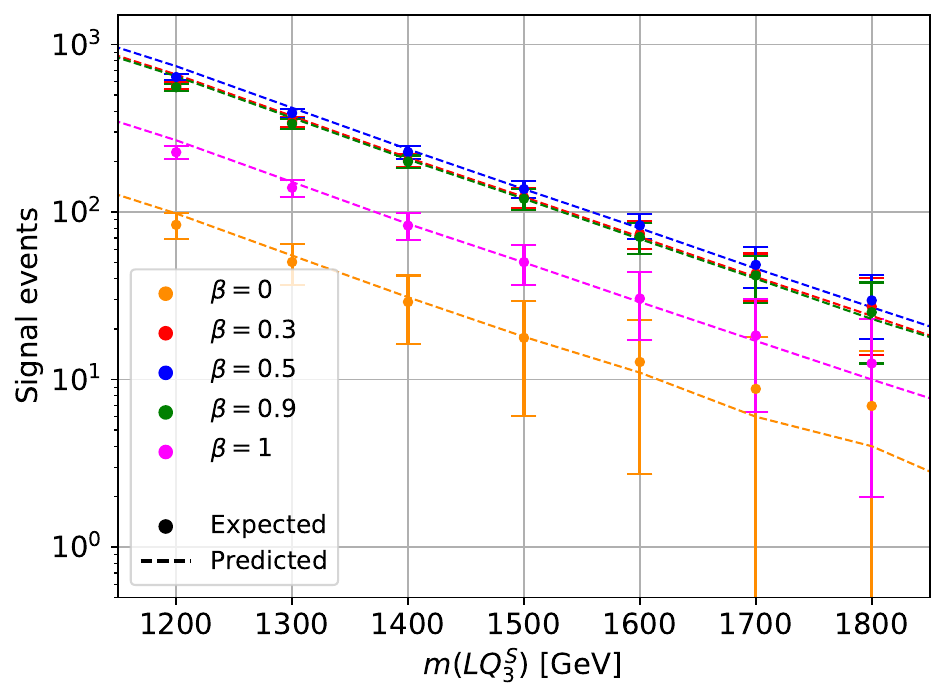}
  \includegraphics[width=0.49\textwidth]{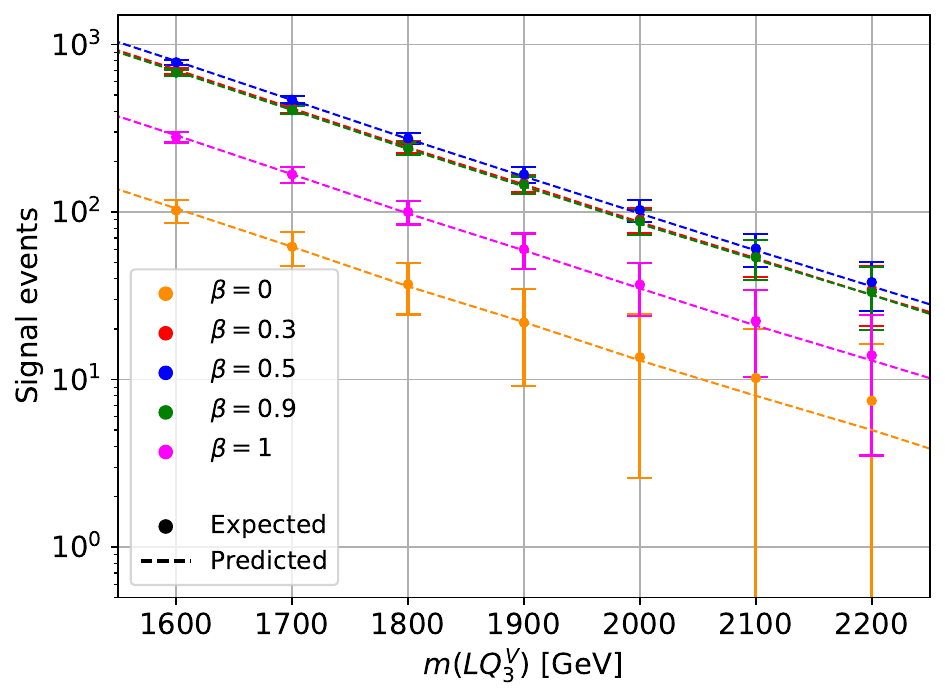}

\caption{Number of predicted (dashed lines) and expected (points) signal events as a function of the mass, for SLQ (left panel) and VLQ (right panel), and different values of $\beta$. The predicted yields, $S_{\text{pred}}$, are obtained multiplying the fiducial cross-section times the luminosity at $\sqrt{s}=14$ TeV and 3000 fb$^{-1}$. The mean expected signal, $\langle S_{\text{exp}}\rangle$, and 1$\sigma$ error bars are computed through a BL fit of the output of Model A over 1000 pseudo-experiments.}
\label{fig:Spredicted}
\end{figure}

The input feature set of Model~B includes all 53 observables listed in Table~\ref{tab:variables}. 
In addition, the total signal yield ($S$) is included as an additional input feature. During the training phase, this quantity can be estimated with the predicted signal yield, $S_{\text{pred}}$, obtained by multiplying the fiducial cross section of each benchmark point by the integrated luminosity, assuming $\sqrt{s}=14$ TeV and 3000 fb$^{-1}$. In order to account for statistical fluctuations, each training event is assigned a value $S$ sampled from a Poisson distribution with parameter $S_{\text{pred}}$. However, for a new pseudo-experiment or a real experimental dataset, the fiducial cross section is not known a priori. Therefore, during the testing and inference phase, we estimate this quantity by performing a BL fit to the Model~A output distribution of the selected events in that pseudo-experiment, using signal and background templates. With this fit we obtain both expected signal and background yields, hereafter denoted as $S_{\text{exp}}$ and $B_{\text{exp}}$, respectively. Finally, all events belonging to the same pseudo-experiment are assigned the same feature value $S = S_{\text{exp}}$ when evaluated by Model~B.

As discussed in Section~\ref{simulation}, scalar and vector leptoquark samples are generated over different mass intervals, reflecting the significantly larger production cross sections of vector leptoquarks. Consequently, scalar and vector benchmark points yielding comparable event rates do not correspond to identical masses, as illustrated in Fig.~\ref{fig:Spredicted}. The inferred signal yield therefore provides valuable information on the range of mass hypotheses compatible with the observed signal and complements the event-level kinematic information used by Model~B. From a machine-learning perspective, the classifier can thus be interpreted as a yield-conditioned model embedded in a hierarchical two-stage inference framework, where the signal strength extracted by Model~A guides the subsequent scalar--vector discrimination.

\begin{figure}
  \centering
  \includegraphics[width=0.5\textwidth]{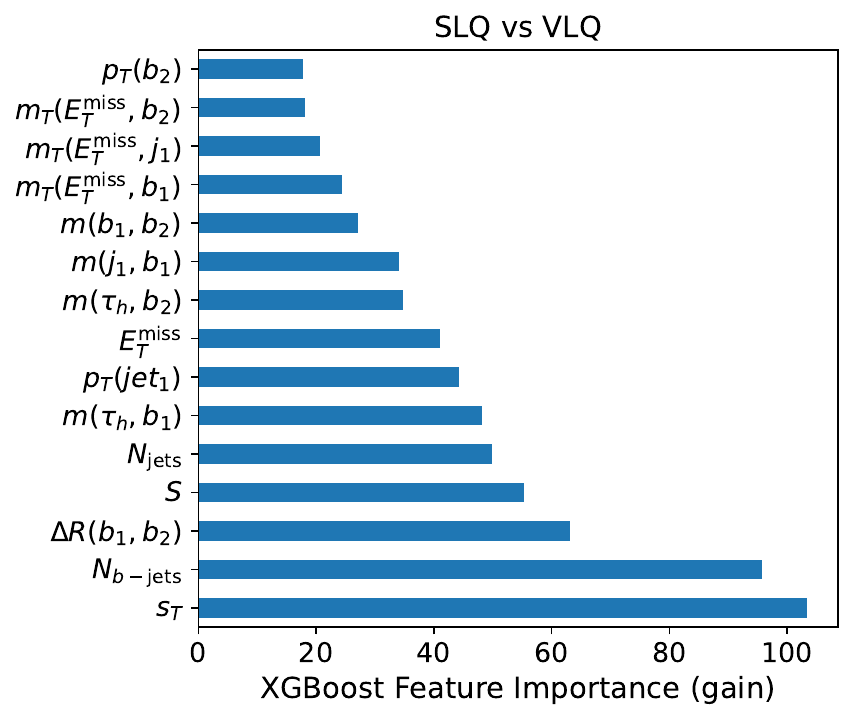}
  \includegraphics[width=0.48\textwidth]{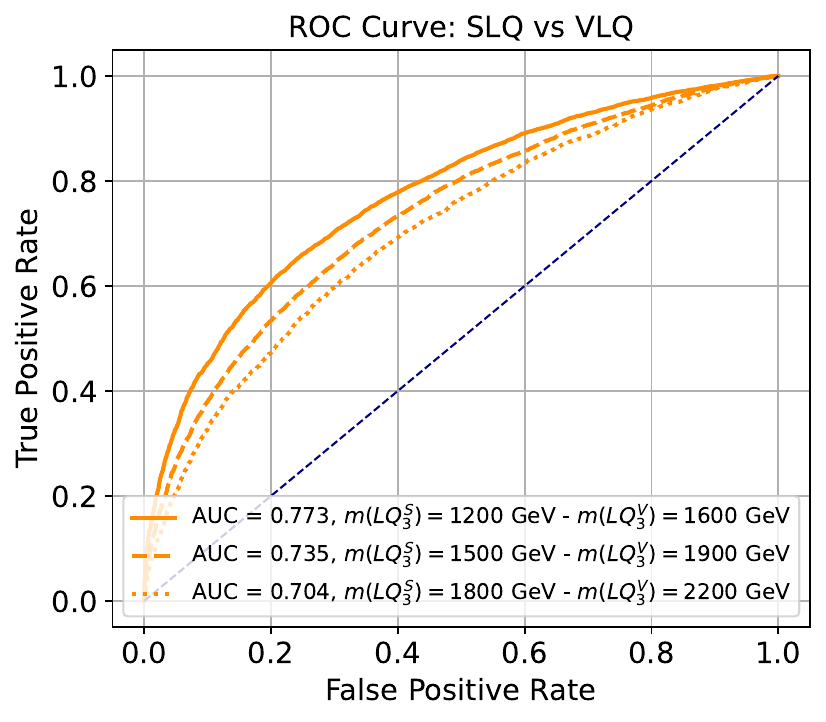}
\caption{Feature-importance ranking for Model B (left panel), computed using the gain metric of the {\tt XGBoost} classifier. The score quantifies the relative contribution of each input variable to the SLQ--VLQ discrimination task. ROC curves for Model B (right panel) for representative SLQ and VLQ benchmark pairs corresponding to alternative spin hypotheses with approximately equal signal yields (within $\sim5\%$).}
\label{XGBoost-outputs-classB}
\end{figure}

Fig.~\ref{XGBoost-outputs-classB} summarizes the performance of Model~B. The left panel shows the feature-importance ranking obtained from the {\tt XGBoost} gain metric, while the right panel displays the corresponding ROC curves evaluated for representative benchmark pairs corresponding to SLQs and VLQs. Each benchmark pair predicts approximately the same signal yield for the two alternative spin hypotheses (equivalently, nearly identical production rates), within a $\sim5\%$ tolerance. The classifier achieves its best performance for the lowest-mass benchmark pair, with an AUC of 0.773 for $m(LQ_3^{S})=1200$ GeV and $m(LQ_3^{V})=1600$ GeV. The discrimination power gradually decreases for higher-mass pairs, reaching an AUC of about 0.701 for $m(LQ_3^{S})=1800$ GeV and $m(LQ_3^{V})=2200$ GeV. This behavior suggests that scalar--vector discrimination becomes increasingly challenging in the high-mass regime as the kinematical distributions broaden and peak at higher energies for higher LQ masses, causing the angular and energy distributions of both spin hypotheses to progressively overlap. The most relevant observables for the scalar--vector discrimination task are found to be $s_T$, the number of $b$-tagged jets, $\Delta R(b_1,b_2)$, and $S$. The prominent role of the latter demonstrates that conditioning the classifier on the signal yield provides additional information beyond the event kinematics alone. This observation supports the interpretation of Model~B as a yield-conditioned classifier and validates the hierarchical inference strategy adopted in this work, where the signal yield extracted by Model~A is used to guide the subsequent spin-discrimination task.

\begin{figure}
  \centering
  \includegraphics[width=0.98\textwidth]{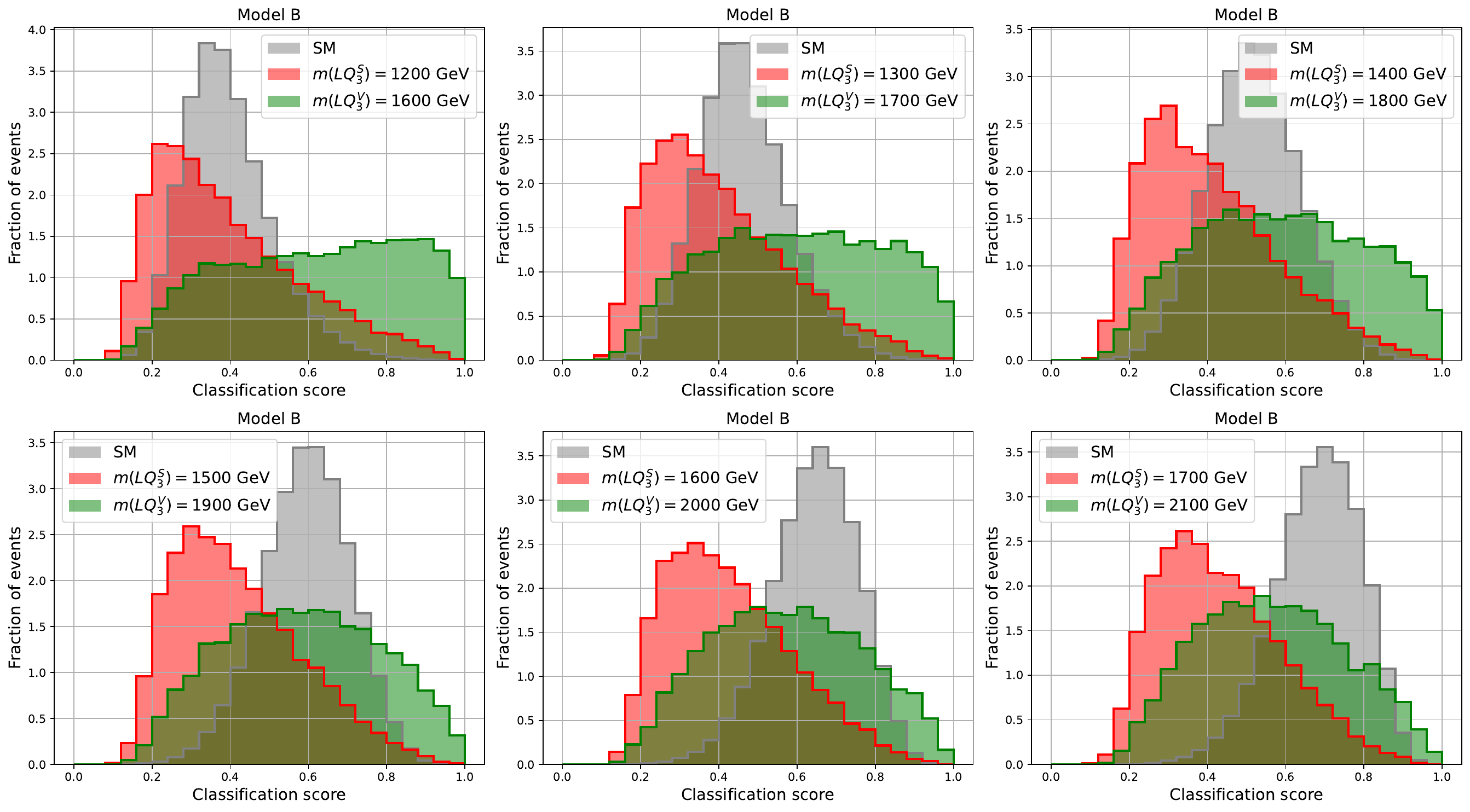}
\caption{
Output distributions of Model~B for representative SLQ (red) and VLQ (green) benchmark pairs in the testing datasets, corresponding to alternative spin hypotheses with approximately equal signal yields (within $\sim5\%$). The grey histogram shows the distribution of SM background events, not used in the training of Model~B, although they represent the residual background contamination that would accompany the selected signal candidates in a realistic experimental analysis.}
\label{fig:MLoutput_classB}
\end{figure}

Fig.~\ref{fig:MLoutput_classB} shows the output distributions of Model~B, for some of the benchmark points in the SLQ and VLQ testing datasets. A stable classifier response is observed throughout the entire mass range considered, with scalar-leptoquark events preferentially populating lower classifier scores and vector-leptoquark events shifted towards larger values. Although the separation gradually decreases towards the higher-mass regime, consistent with the reduction in classification performance observed in the ROC curves, the ordering of the two hypotheses remains unchanged. 

For completeness, each panel in Fig.~\ref{fig:MLoutput_classB} also includes the distribution of SM background events. These events were not used during the training of Model~B, which is trained exclusively to distinguish between scalar and vector leptoquark hypotheses. The SM background populates a broad range of classifier scores, with its distribution centered around intermediate values. While this introduces uncertainty, as SM events may be confused with either scalar or vector signatures depending on their masses, its impact on the final analysis is not significant. This is because Model B is applied after selecting a signal-enriched region defined by a threshold on the Model A classifier score. Given that Model A provides an excellent separation between the SM background and LQ signals (AUC$=0.997$), the background contamination is low. Nonetheless, this uncertainty source is taken into account in the statistical procedure and final significance calculations via pseudo-experiments that incorporate both signal and background events.

Finally, we define a test statistic that provides a global measure of the significance of Model B in discriminating between scalar and vector leptoquarks. This statistic is constructed as the sum over all events that pass the working point ($\mathrm{WP}$) defined by Model A in a given pseudo-experiment ($N_{\mathrm{WP}}=\{i \in \{1,...,N_{\mathrm{pseudo}}\} | o_A(x_i)>\mathrm{WP}\}$), of the logarithm of the likelihood ratio between the probabilities of each event being classified as scalar-like or vector-like:
\begin{equation}
\label{eq:test}
    T = \sum_{i \in N_{\mathrm{WP}}}
    \log\left(\frac{p_{\mathrm{vector},i}}{p_{\mathrm{scalar},i}}\right)
    =
    \sum_{i \in N_{\mathrm{WP}}}
    \log\left(\frac{o_B(x_i)}{1-o_B(x_i)}\right),
\end{equation}
where $o_B(x)\in[0,1]$ is the output of the Model B classifier, with values close to $o_B(x)=0$ corresponding to SLQ-like events and $o_B(x)=1$ to VLQ-like events.

\begin{figure}
  \centering
  \includegraphics[width=0.5\textwidth]{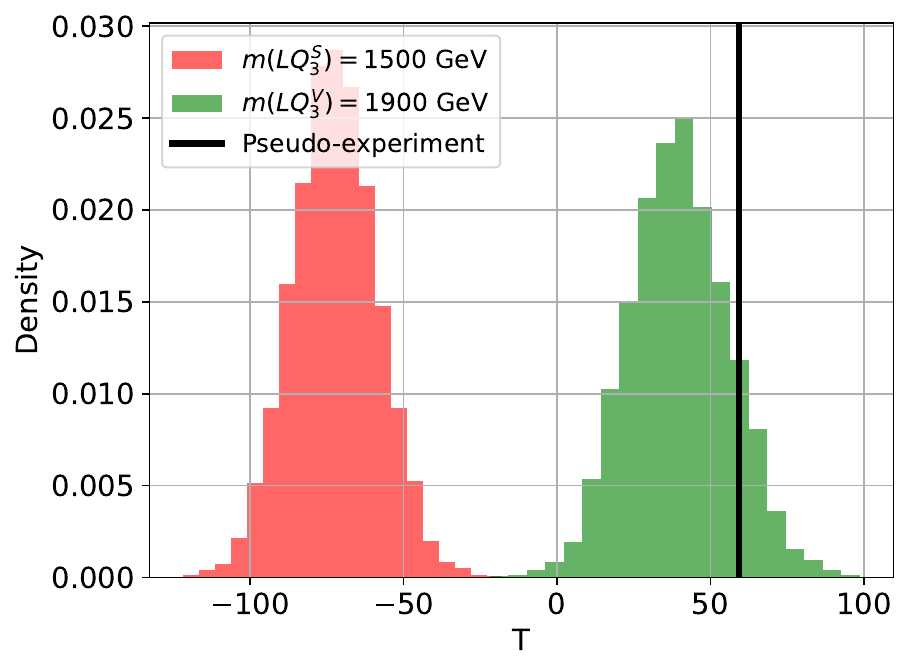}
\caption{Distribution of the $T$-statistic obtained from $1000$ pseudo-experiments for two representative benchmark points: a SLQ with $m(LQ_3^{S})=1500~\mathrm{GeV}$ (red) and a VLQ with $m(LQ_3^{V})=1900~\mathrm{GeV}$ (green). The corresponding Gaussian fits yield $(\mu_s,\sigma_s)=(-71.3,\,14.1)$ and $(\mu_v,\sigma_v)=(40.0,\,16.5)$, respectively. The vertical solid black line indicates the value of the test statistic $T$ obtained for a single pseudo-experiment, representing the measurement that would be obtained from one experimental realization.}
\label{fig:example-stat-T_AUX}
\end{figure}

\begin{figure}
  \centering
  \includegraphics[width=0.5\textwidth]{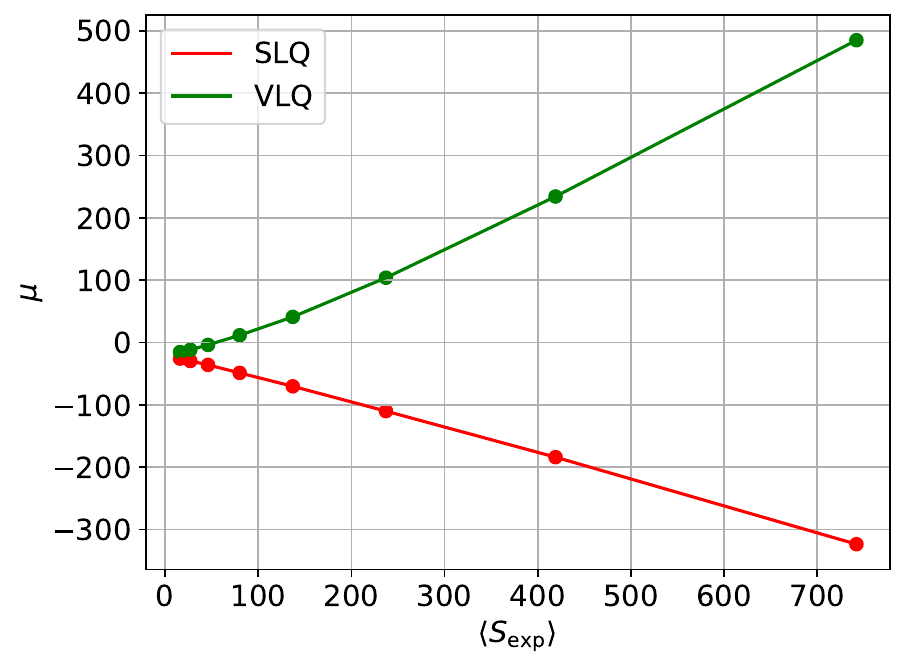}
  \includegraphics[width=0.48\textwidth]{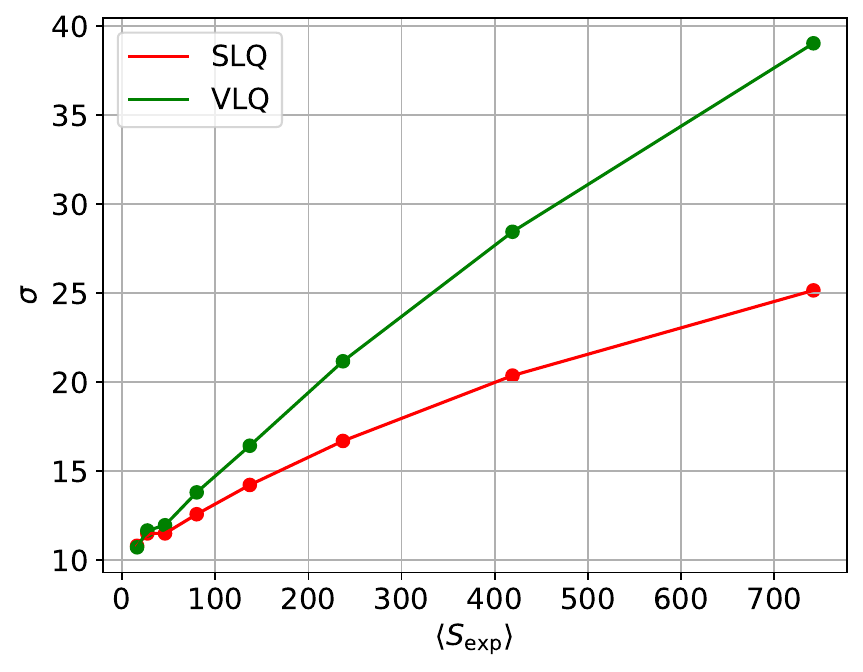}
  \caption{
Dependence of the Gaussian parameters $\mu$ (left panel) and $\sigma$ (right panel), extracted from Gaussian fits to the $T$-statistic distributions, on the mean expected signal yield, $\langle S_{\mathrm{exp}}\rangle$, for the SLQ (red) and VLQ (green) hypotheses. The points correspond to the simulated benchmark scenarios, while the solid lines show the linear fits used to interpolate the Gaussian parameters for intermediate signal yields.}
\label{fig:mu_sigma_S}
\end{figure}

To calibrate the $T$-statistic, we generate $1000$ pseudo-experiments for each benchmark point. Each pseudo-experiment is composed of signal and background events independently drawn from Poisson distributions with mean values equal to the predicted number of signal ($S_{\text{pred}}$) and background ($B_{\text{pred}}$) events, respectively. Then, we compute the test statistic defined in Eq.~(\ref{eq:test}) for each pseudo-experiment. As an illustration, Fig.~\ref{fig:example-stat-T_AUX} shows the distribution of the $T$-statistic for two representative benchmark points with approximately equal signal yields: a scalar leptoquark with $m(LQ_3^{S})=1500~\mathrm{GeV}$ (red) and a vector leptoquark with $m(LQ_3^{V})=1900~\mathrm{GeV}$ (green). In both cases, the resulting distributions are well described by Gaussian functions. The same behavior is observed for all benchmark points considered, although with different values of the corresponding means and standard deviations. The dependence of the Gaussian parameters, $\mu_{S,V}$ and $\sigma_{S,V}$ (where the subscripts $S$ and $V$ denote the scalar and vector hypotheses, respectively), on the mean inferred signal yield, $\langle S_{\text{exp}} \rangle$, is shown in Fig.~\ref{fig:mu_sigma_S}. In order to interpolate between the simulated benchmark points, we perform linear fits to the $\mu_{S,V}(S_{\text{exp}})$ and $\sigma_{S,V}(S_{\text{exp}})$ values. This procedure allows us to estimate the Gaussian parameters for any intermediate signal yield, and therefore for any scalar or vector leptoquark mass within the ranges $m(LQ_3^{S})\in[1200,1900]~\mathrm{GeV}$ and $m(LQ_3^{V})\in[1600,2300]~\mathrm{GeV}$, respectively. For example, for $\sqrt{s}=14$~TeV and an integrated luminosity of $3000~\mathrm{fb}^{-1}$, these mass ranges correspond to expected signal yields in the interval $S_{\text{exp}}\in[742,16]$. For a given pseudo-experiment (or, eventually, real experimental data) we compute the corresponding value of the test statistic $T$ (represented as a  vertical solid line in the example shown in Fig.~\ref{fig:example-stat-T_AUX}). We then quantify its compatibility with the scalar and vector hypotheses in terms of the corresponding significance:
\begin{equation}
\label{eq:z}
    \mathcal{Z}_{S,V}(S_{\text{exp}})=
    \frac{|T-\mu_{S,V}(S_{\text{exp}})|}{\sigma_{S,V}(S_{\text{exp}})}.
\end{equation}

\begin{figure}
  \centering
  \includegraphics[width=0.49\textwidth]{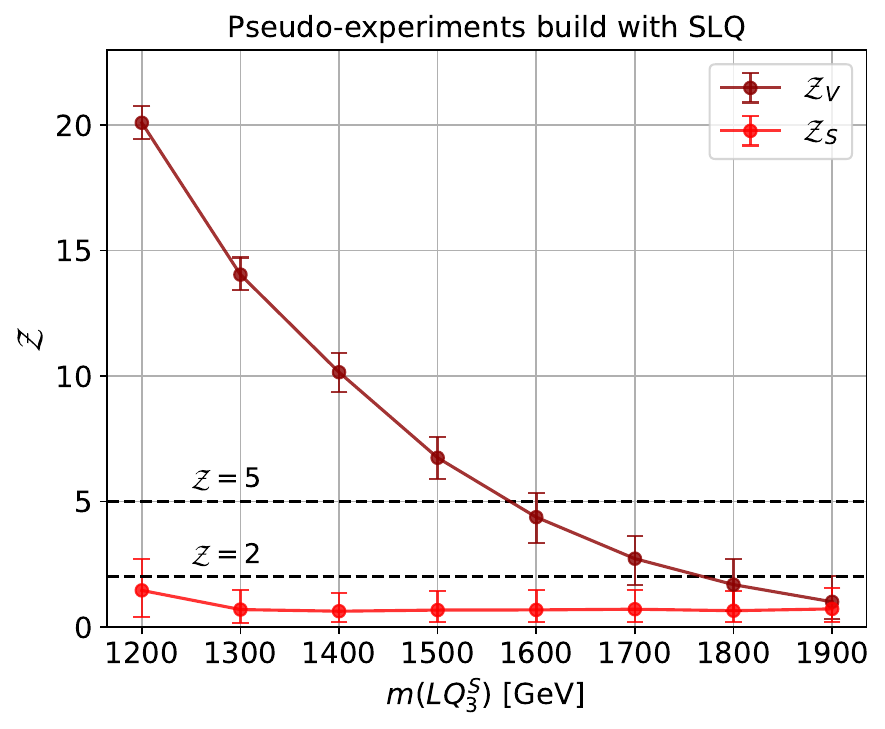}
  \includegraphics[width=0.49\textwidth]{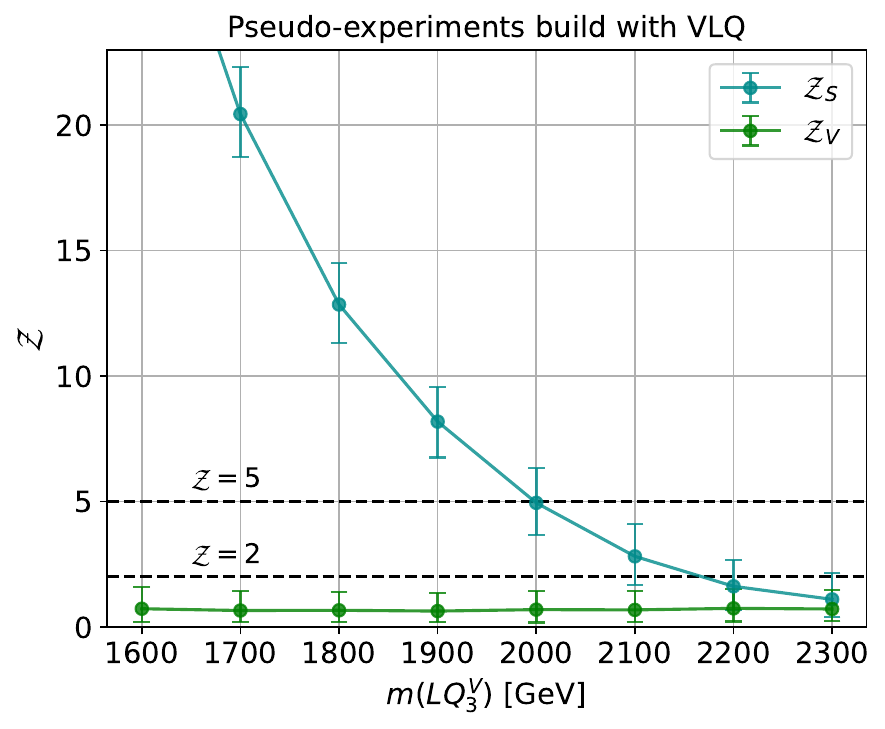}
  \caption{
Significance results with Model B for the considered SLQ (VLQ) benchmark points as a function of $m(LQ_3^{S})$ ($m(LQ_3^{V})$) in the left (right) panel. For each benchmark point, the marker indicates the mean significance obtained from 1000 pseudo-experiments, while the error bars represent the $1\sigma$ spread of the corresponding significance distribution. The continuous curves show the interpolation between the simulated benchmark points.}
\label{fig:signif}
\end{figure}

The corresponding distributions of the significances $\mathcal{Z}_S$ and $\mathcal{Z}_V$ over the considered  mass ranges are shown in Fig.~\ref{fig:signif}. We computed the value of $\mathcal{Z}_S$ and $\mathcal{Z}_V$ for 1000 pseudo-experiments. The markers indicate the mean significance, while the error bars represent the $1\sigma$ spread of the significance distribution. Adopting a conservative criterion, we define a pseudo-experiment to be compatible with the scalar (vector) hypothesis if $\mathcal{Z}_{S}<2$ ($\mathcal{Z}_{V}<2$) and incompatible with the alternative hypothesis if $\mathcal{Z}_{V}>5$ ($\mathcal{Z}_{S}>5$). A pseudo-experiment is identified as scalar (vector) only when both conditions are simultaneously satisfied. It is important to emphasize that these two conditions are not complementary, i.e., a pseudo-experiment may be compatible with both hypotheses, or with neither of them. In such cases, no unambiguous statement can be made regarding the scalar or vector nature of the observed signal. In Appendix~\ref{sec:benchmarksVAL} we present the application of the procedure using four benchmark points. In particular, we show that the procedure is robust and can be implemented for datasets generated with masses that were not used during training or testing.

To ensure reproducibility, the full analysis framework, including data-processing routines, model-training scripts, and statistical-analysis tools, is publicly available in~\cite{SLQvsVLQ-code}.

\section{Results} \label{Results}

In this section we present the main results obtained with the two-stage inference framework introduced in the previous sections. We first derive the expected discovery and exclusion regions in the $(m(LQ_3^{S/V}),\beta)$ plane, with $\beta={\rm BR}(LQ_3^{S/V}\rightarrow b\tau)$, using the statistical analysis based on the output of Model~A. We then assess the performance of Model~B in discriminating between scalar and vector leptoquark hypotheses, identifying the regions of parameter space where the spin nature of a potential signal can be reliably established.

\begin{figure}
  \centering
  \includegraphics[width=0.85\textwidth]{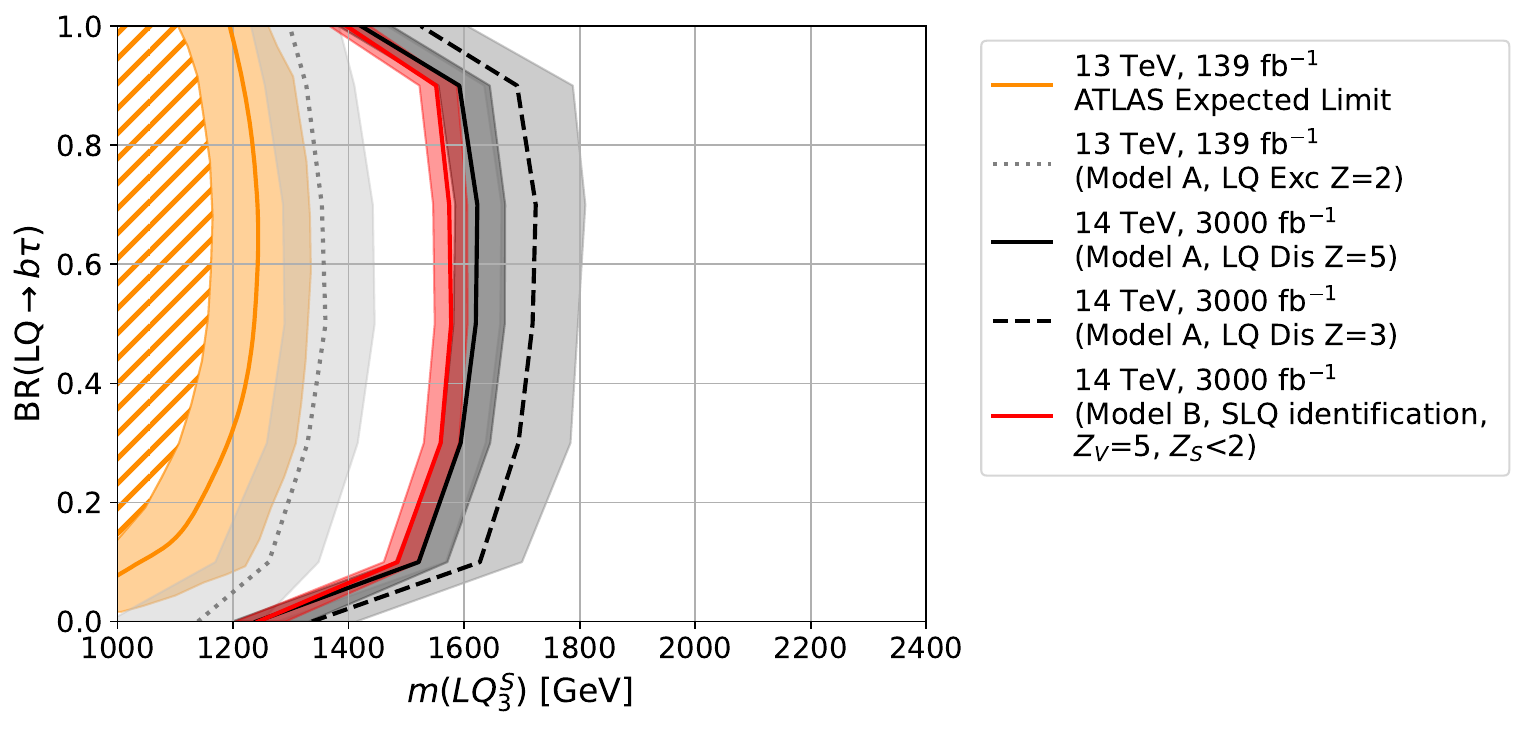}\\
  \includegraphics[width=0.85\textwidth]{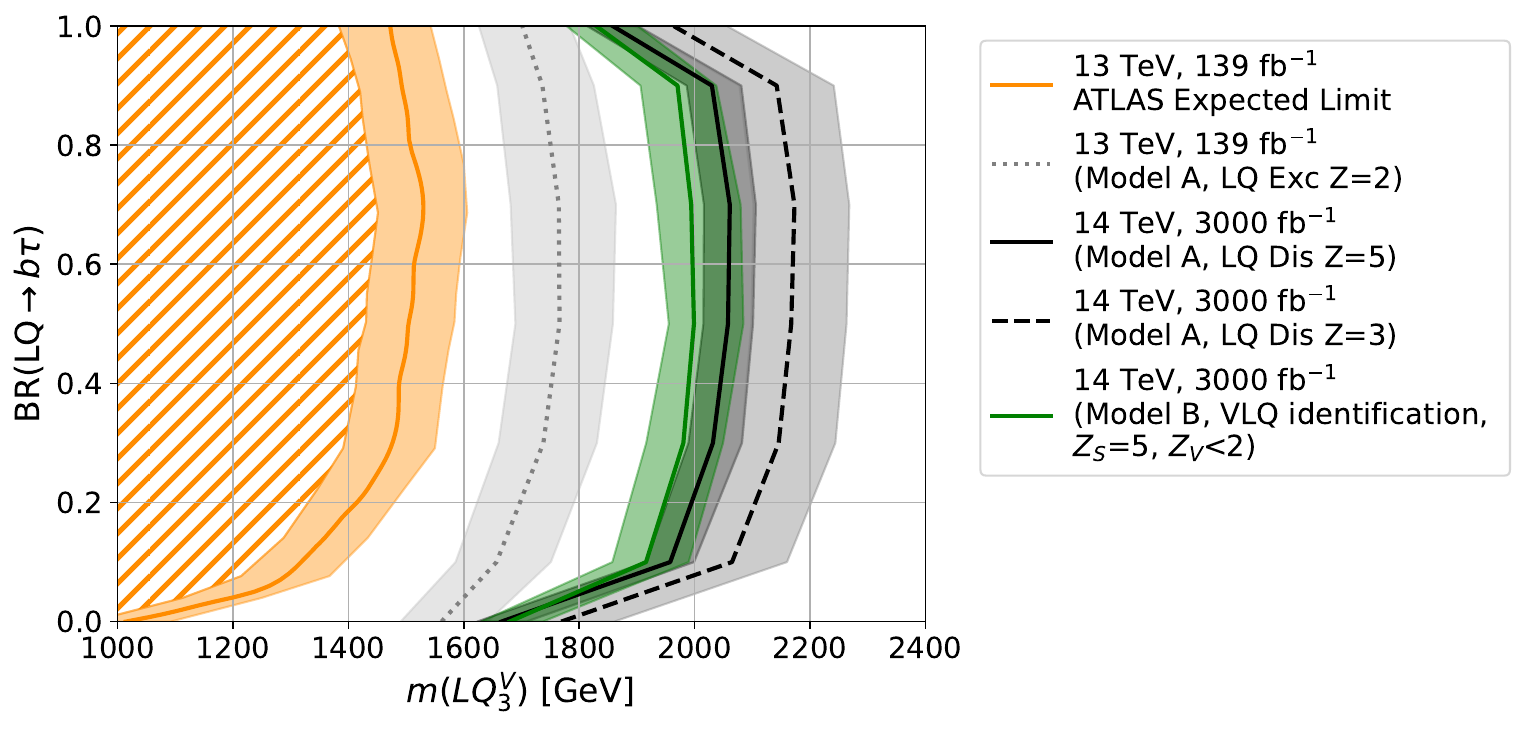}
\caption{
Expected HL-LHC discovery ($5\sigma$, black solid lines) and evidence ($3\sigma$, black dashed lines) contours for the SLQ (upper panel) and VLQ (lower panel) hypotheses in the plane of the leptoquark mass, $m(LQ_3^{S/V})$ and the branching fraction $\mathrm{BR}(LQ_3^{S/V}\rightarrow b\tau)$. The contours are obtained from the BL analysis using Model~A  with the corresponding discovery statistical tests. The orange region is excluded by current experimental searches using the full LHC Run~2 dataset~\cite{ATLAS:2021jyv}. For comparison, the expected LHC Run~2 exclusion contour ($2\sigma$, light-grey dotted lines), obtained from Model~A, is also shown. The red (upper panel) and green (lower panel) solid contours delimit the regions where Model~B provides sufficient discrimination power to identify the observed signal as originating from a SLQ ($\mathcal{Z}_V>5$ and $\mathcal{Z}_S<2$) or a VLQ ($\mathcal{Z}_S>5$ and $\mathcal{Z}_V<2$), according to the criterion defined in Sec.~\ref{modelB}. In all cases, the associated $\pm1\sigma$ statistical uncertainty contours are also shown; no systematic uncertainties are included.
}
\label{fig:discovery_exclusion_limits}
\end{figure}

In the upper (lower) panel of Fig.~\ref{fig:discovery_exclusion_limits} we present the main results obtained for the scalar (vector) leptoquark scenario in the $(m(LQ_3^{S/V}),\beta)$ plane. We first discuss the results obtained with Model~A, which is responsible for the discrimination between leptoquark signals and the SM background.

The light-grey dotted contours correspond to the expected LHC Run~2 exclusion limits obtained with our ML-based analysis using the BL exclusion test. For intermediate values of the branching fraction, these contours extend up to approximately $m(LQ_3^{S})\simeq1.344$~TeV and $m(LQ_3^{V})\simeq1.748$~TeV for scalar and vector leptoquarks, respectively. As $\beta$ approaches either 0 or 1, the fraction of signal events containing exactly one hadronically decaying tau lepton decreases, reducing the signal acceptance and consequently weakening the expected exclusion reach. Overall, the multivariate analysis exhibits a clear tendency towards improving the exclusion limits with respect to those currently achieved by ATLAS Collaboration~\cite{ATLAS:2021jyv}, a behavior already observed in our previous study~\cite{Arganda:2023qni}. Furthermore, the expected Run~2 sensitivities obtained with the present implementation of Model~A surpass those reported in Ref.~\cite{Arganda:2023qni}. As discussed in Sec.~\ref{modelA}, this improvement can be attributed primarily to the significantly richer set of input observables employed in the present work, which allows the classifier to exploit additional kinematic information beyond that available in the previous analysis. Although an unbinned statistical treatment would likely provide a further increase in sensitivity, its considerably higher computational cost is not justified for the objectives of the present work, where Model~A is primarily intended to identify signal-like events and provide the signal-yield estimate required by the second stage of the inference framework. For completeness, the full set of ATLAS exclusion limits obtained from searches targeting different final states, together with our corresponding exclusion curves for Run~2, Run~3, and the HL-LHC, are presented in Appendix~\ref{app:ML}.

The black dashed contours of Fig.~\ref{fig:discovery_exclusion_limits} show the expected HL-LHC $3\sigma$ evidence reach obtained with the BL discovery test, assuming $\sqrt{s}=14$~TeV and an integrated luminosity of $3000~\mathrm{fb}^{-1}$, defining the region of parameter space where evidence for new physics could be established. More importantly, the black solid contours correspond to the expected $5\sigma$ discovery reach derived from the same BL discovery test. The latter constitutes the most relevant result of Model~A, as they delimit the region of parameter space in which a leptoquark signal could potentially be discovered with future HL-LHC data. Once such a signal is established, the subsequent challenge is no longer its observation, but rather the determination of its underlying nature. Consequently, the parameter region enclosed by the discovery contours defines precisely the domain in which the spin-discrimination strategy developed with Model~B becomes essential.

We now turn to the second stage of the inference framework, namely the spin discrimination performed by Model~B. The red (green) solid contours in Fig.~\ref{fig:discovery_exclusion_limits} delimit the regions in which a scalar (vector) leptoquark signal can not only be identified as such, but also shown to be incompatible with the alternative vector (scalar) hypothesis according to the statistical procedure introduced in Sec.~\ref{modelB}~\footnote{While both criteria are simultaneously applied to classify individual pseudo-experiments, the parameter region in Fig.~\ref{fig:discovery_exclusion_limits} where the scalar (vector) nature can be identified is determined by the $\mathcal{Z}_{V}>5$ ($\mathcal{Z}_{S}>5$) contour. Since the signal compatibility condition is satisfied on average throughout the mass range (see Fig.~\ref{fig:signif}), the exclusion of the alternative hypothesis sets the primary constraint on the identification limit, failing only due to statistical fluctuations.}. The corresponding shaded bands indicate the associated $\pm1\sigma$ statistical uncertainties on these contours.

The most important outcome of the present work is that, even after accounting for these statistical uncertainties, the spin-discrimination reach closely follows the discovery potential obtained with Model~A. In other words, throughout essentially the entire region of parameter space where a leptoquark signal could be discovered with a significance exceeding $5\sigma$, our framework is also able to determine whether the observed resonance is compatible with the scalar or vector hypothesis and incompatible with the
alternative one. Therefore, the proposed two-stage strategy not only provides the discovery potential for third-generation leptoquarks, but also delivers a reliable characterization of the underlying particle once a signal is established.

This result is particularly remarkable given the deliberately conservative criterion adopted in Sec.~\ref{modelB} to claim compatibility with one spin hypothesis while excluding the alternative. Our procedure requires a robust separation between the corresponding classifier-response distributions before assigning the spin of the observed signal. Consequently, the discrimination regions shown in Fig.~\ref{fig:discovery_exclusion_limits} should be regarded as conservative estimates of the actual spin-identification capability of the method.

\subsection{Dependence on the Model A working point and $\boldsymbol{\beta}$ parameter}
\label{sec:threshold_dependence}

\begin{figure}
  \centering
  \includegraphics[width=0.49\textwidth]{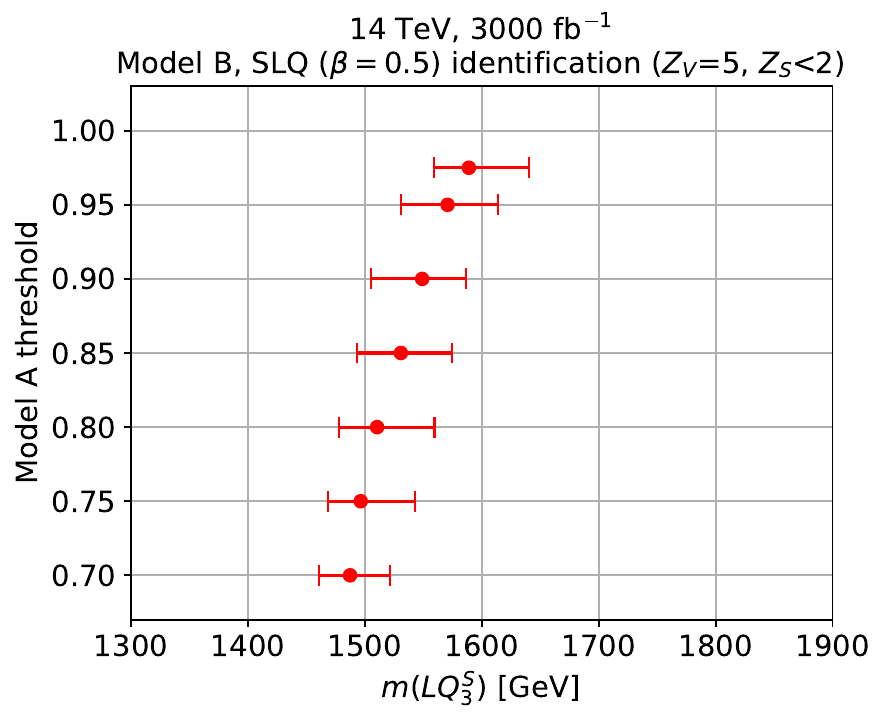}
  \includegraphics[width=0.49\textwidth]{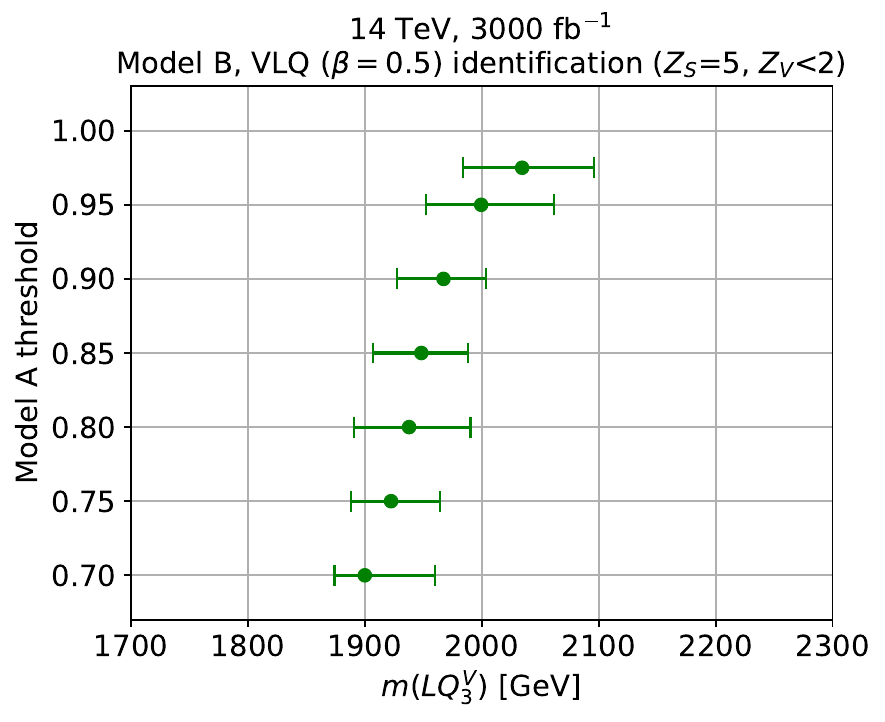}
\caption{
Dependence of the Model~B mass reach on the threshold applied to the output of Model~A for $\sqrt{s}=14$~TeV and an integrated luminosity of 3000~fb$^{-1}$. The left (right) panel corresponds to pseudo-experiments containing SLQ (VLQ) events, with $\beta=0.5$. The mass reach is defined by the compatibility requirements $Z_V=5$, $Z_S<2$ ($Z_S=5$, $Z_V<2$) for the SLQ (VLQ) hypothesis.
}
\label{fig:thresh}
\end{figure}

As discussed in Sec.~\ref{modelA}, we adopt a reference working point of 0.95 for the Model A classifier output when defining the signal-enriched region used as input to Model B. To assess the dependence of the Model B results on this choice, we repeat the analysis for a range of Model A thresholds. The corresponding results are shown in Fig.~\ref{fig:thresh} for the SLQ and VLQ hypotheses at $\sqrt{s}=14$~TeV with an integrated luminosity of 3000~fb$^{-1}$, fixing $\beta=0.5$.

For the SLQ hypothesis, the maximum mass compatible with the requirements $Z_V=5$ and $Z_S<2$ increases gradually from about $1.5$~TeV at a threshold of 0.70 to about $1.6$~TeV for the largest thresholds considered. Similarly, for the VLQ hypothesis, imposing $Z_S=5$ and $Z_V<2$, the corresponding mass reach increases from about $1.9$~TeV to about $2.0$~TeV. In both cases, the dependence on the Model A threshold is smooth, with no significant change in the overall sensitivity over the range considered.

In particular, the reference working point of 0.95 lies close to the region where the sensitivity has already reached its maximal values, while providing a suitable compromise between retaining a statistically robust signal sample and suppressing the residual SM background contamination. While this choice does not correspond to the point of maximum sensitivity in the scan, it provides a sufficiently pure and statistically populated sample to ensure a robust SLQ--VLQ discrimination with Model~B throughout the region where Model~A provides discovery sensitivity. 

In addition, we evaluated the impact of the branching fraction parameter $\beta$ on the discrimination power. Throughout this work, the computation of the significance sensitivity using Model A and Model B, along with the determination of the parameters $\mu_{S,V}$ and $\sigma_{S,V}$ describing the $T$ statistic distributions, assume that the value of $\beta$ is known, given that these quantities directly depend on the signal yield. However, for a given LQ mass, the overall number of predicted events remains virtually flat across the range $\beta \in [0.3, 0.9]$ (see Fig.~\ref{fig:Spredicted}). Then, the sensitivity of both Models is expected to be largely insensitive to the exact choice of $\beta$. We verified that applying the parameterizations set with $\beta=0.5$ (the final state channel and selection cuts were optimized for this scenario) across $\beta \in [0.3, 0.9]$ leads to negligible variations in the signal identification reach, as well as in the exclusion and discovery limit. Consequently, the presented strategy demonstrates remarkable robustness and stable scalar--vector discrimination independently of the Model A working point, and the specific value of $\beta$ (within $[0.3, 0.9]$).

\subsection{Approach to the Inclusion of Systematic Uncertainties}
\label{systematics}

A realistic assessment of the proposed analysis requires estimating the impact of systematic uncertainties on the statistical inference. Since a complete experimental treatment based on nuisance parameters is beyond the scope of this work, we adopt a simplified procedure aimed at evaluating the robustness of the machine-learning strategy against variations of the most relevant input observables. This approach provides a first estimate of the stability of the expected sensitivities while keeping the analysis at the level of the reconstructed kinematic variables used to train the ML classifiers.

Instead of propagating detector and theoretical uncertainties through the full simulation chain, we evaluate their effect directly in the feature space. In particular, we consider systematic variations of the input observables entering the ML models, neglecting correlations among them. As shown by the feature-importance rankings in Figs.~\ref{XGBoost-outputs} and~\ref{XGBoost-outputs-classB}, the variable $s_T$ provides the largest contribution to the discrimination power of both Model~A and Model~B. We therefore use it as a representative observable to estimate the impact of systematic effects.

Following the strategy adopted in Refs.~\cite{Arganda:2023qni, Arratia:2021otl,CMS:2022ytw}, we evaluate the trained classifiers on modified test samples in which all event features remain unchanged except for the selected observable. More specifically, for each event we replace
\begin{equation}
s_T \rightarrow s_T(1\pm\delta),
\end{equation}
with $\delta=0.10$, corresponding to a conservative $10\%$ variation, compatible with the typical size of experimental uncertainties affecting global kinematic observables~\cite{ATLAS:2017mpa}. The ML models are not retrained; only the evaluation samples are modified. This procedure produces two alternative classifier responses, denoted by $o(x)^+$ and $o(x)^-$, corresponding to the upward and downward variations, respectively. To obtain a conservative estimate of the sensitivity, we quote the least restrictive result obtained among the different systematic variations considered.

\begin{figure}
  \centering
  \includegraphics[width=0.85\textwidth]{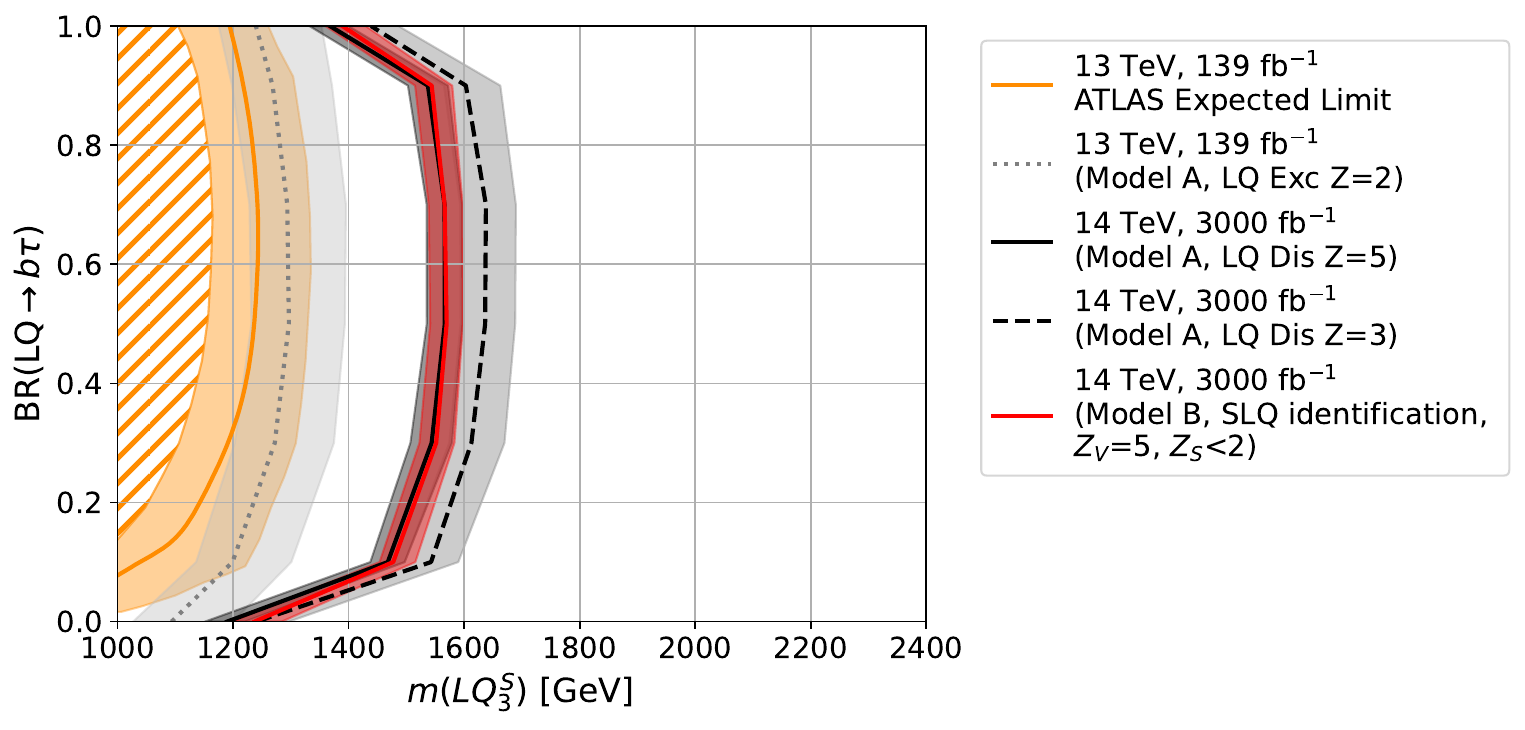}\\
  \includegraphics[width=0.85\textwidth]{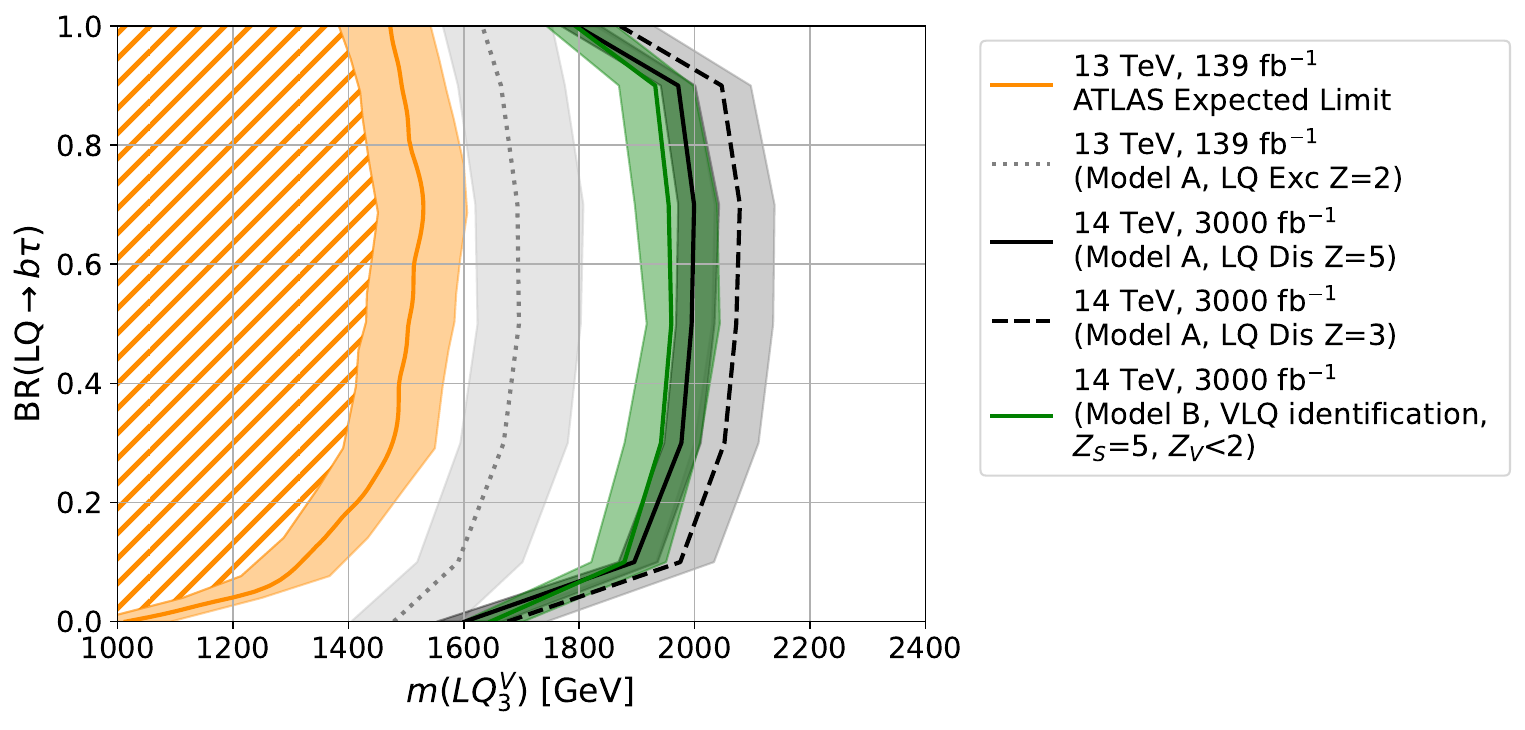}
\caption{
Same as Fig.~\ref{fig:discovery_exclusion_limits}, but including the effect of the systematic uncertainties discussed in Sec.~\ref{systematics}. The contours are obtained after propagating a conservative $\pm10\%$ variation of the $s_T$ observable through the ML-based statistical analysis, providing an estimate of the robustness of the expected discovery, exclusion, and SLQ--VLQ discrimination regions under systematic variations of the most relevant input feature.
}
\label{LQ-withsys}
\end{figure}

The corresponding results are presented in Fig.~\ref{LQ-withsys}. Compared with the nominal analysis shown in Fig.~\ref{fig:discovery_exclusion_limits}, the discovery and exclusion contours remain largely stable over most of the parameter space, changing only at the level of a few percent. Crucially, this stability holds in the discovery region with more than $5\sigma$ at the HL-LHC, where spin identification is performed.  A more noticeable effect is observed in the highest-mass region, where the separation between the $5\sigma$ and $3\sigma$ contours for the HL-LHC scenario becomes more distorted than the separation between the $5\sigma$ discovery contour and the $2\sigma$ exclusion contour for the full Run~2 LHC dataset. This behaviour occurs in the regime of very small expected signal yields, where the statistical inference is naturally more sensitive to changes in the classifier response. In addition, as shown in Fig.~\ref{distributions}, the $s_T$ spectrum becomes significantly harder for the largest leptoquark masses. In this regime, applying a uniform $\pm10\%$ rescaling of $s_T$ constitutes an excessively conservative approximation, since it induces distortions of the feature distribution that are unlikely to be representative of realistic experimental systematic effects. It is worth recalling that this is a simplified treatment of systematic uncertainties, which may become inaccurate in the low-signal regime. Therefore, the shift observed in the relative separation between the contours at the highest masses should be interpreted with caution, rather than as a significant physical effect. We have also repeated the same procedure for other relevant input observables and found similarly small effects on the final results.

It is important to emphasize that this study should be regarded only as a first assessment of the robustness of the proposed ML framework. A complete experimental analysis would require incorporating all relevant detector and theoretical systematic uncertainties, together with their associated nuisance parameters and correlations, within the likelihood fit. Nevertheless, the modest dependence observed under the variations considered here suggests that the proposed strategy is relatively stable against moderate distortions of its most influential input features.

\section{Conclusions}
\label{Conclu}

The observation of a leptoquark signal at the LHC would immediately open a second and equally important question beyond discovery itself: determining the nature of the new state. In particular, establishing whether the resonance corresponds to a scalar or a vector leptoquark would provide essential information about the underlying ultraviolet theory and the dynamics responsible for its interactions. Motivated by this challenge, we have developed a dedicated machine-learning framework designed not only to identify a leptoquark signal but also to characterize its spin in a statistically consistent way.

The proposed strategy relies on a hierarchical two-stage inference pipeline. In the first stage, a supervised classifier is trained to separate leptoquark events from the SM background. Its output is exploited both to define a highly signal-enriched event sample and to infer the signal yield through a binned-likelihood fit analysis. In the second stage, a dedicated classifier distinguishes between SLQ and VLQ hypotheses using the kinematic properties of the selected events together with the signal yield inferred in the first stage. This sequential workflow closely mirrors the procedure that would naturally follow a potential discovery at the LHC, where establishing the presence of a signal necessarily precedes its detailed characterization.

The framework has been investigated for third-generation leptoquarks decaying into final states containing a hadronically decaying tau lepton, $b$-tagged jets, and missing transverse momentum. Scalar and vector benchmark scenarios were considered over a broad mass range, implicitly pairing mass hypotheses with comparable signal yields in order to isolate genuine spin-dependent effects from the trivial differences associated with the larger production cross sections of vector leptoquarks.

Our results demonstrate that the first-stage classifier provides an excellent separation between signal and background events, achieving an AUC close to unity and allowing the definition of highly pure signal samples suitable for subsequent analyses. These results not only confirm the robustness of the strategy introduced in Ref.~\cite{Arganda:2023qni}, but also surpass its performance thanks to the significantly expanded set of input observables considered in the present work, which includes additional high-level variables sensitive to event topology and characteristic mass scales. More importantly, the second-stage classifier successfully exploits the remaining differences in event topology and kinematics to discriminate between scalar and vector hypotheses. We find that incorporating the signal yield inferred from the first-stage likelihood fit significantly improves the classifier performance by providing complementary information on the mass scale compatible with the observed signal. This illustrates the advantage of combining event-level information with global statistical inference within a unified machine-learning framework.

The most important outcome of the present work is that the spin-discrimination reach closely follows the discovery potential obtained with the first-stage analysis. Throughout essentially the entire region of parameter space where a third-generation leptoquark could be discovered with a significance exceeding the conventional $5\sigma$ threshold, the proposed framework is also capable of determining whether the observed resonance can be identified with the scalar or vector hypothesis with comparable statistical significance. This result demonstrates that determining the spin of a newly discovered leptoquark does not necessarily require substantially larger datasets than those needed for discovery itself. Instead, particle characterization can naturally accompany the discovery process.

We have also performed a first assessment of the impact of systematic uncertainties on the proposed strategy. Although the present study is not intended to provide a complete experimental treatment of systematic effects, the results indicate that the overall performance of the framework remains robust under realistic variations of the dominant uncertainties. This provides encouraging evidence that the proposed methodology retains its discriminating power beyond the idealized statistical scenario considered in the baseline analysis, although a more comprehensive treatment incorporating the full experimental uncertainty model will be required in future studies.

While the present analysis focuses on third-generation leptoquarks, the methodology itself is considerably more general. The two-stage inference strategy is not tied to the specific flavour structure or production mechanism considered here and could be extended to alternative leptoquark scenarios, different decay channels, or, more broadly, to new-physics searches involving competing signal hypotheses. Future developments may also explore more sophisticated machine-learning architectures, additional inference techniques, and a full treatment of systematic uncertainties under High-Luminosity LHC conditions.

In summary, we have introduced a statistically consistent machine-learning framework that connects signal discovery and particle characterization within a unified analysis strategy. Beyond providing excellent sensitivity to the presence of a leptoquark signal, the proposed method demonstrates that the spin of a newly discovered resonance can be determined with high significance over essentially the same region of parameter space where discovery is possible. We therefore believe that this approach provides a promising avenue for future collider analyses, where establishing the existence of new particles and unveiling their fundamental properties should be regarded as complementary objectives of a single inference framework.

\vspace{2.5mm}
\paragraph{\small Acknowledgments.}
%
%
{\small
This publication has been funded within the framework of the R\&D\&I Project CEX2025-001574-S, funded by MICIU/AEI/10.13039/501100011033 (EA, RMSS). The research presented in this publication falls within the research line Particle Physics in the Standard Model and Beyond (EA, RMSS). EA and RMSS also acknowledge partial financial support by the Spanish Research Agency (Agencia Estatal de Investigaci\'on) through the Grants IFT Centro de Excelencia Severo Ochoa No CEX2020-001007-S, PID2021-124704NB-I00, and PID2025-172338NB-I00 funded by MCIN/AEI/10.13039/501100011033. MdlR acknowledge financial support from CONICET (PIP 11220210100064CO), Argentina. ADP is supported by a Simons Foundation’s fellowship through the Targeted Grant to Instituto Balseiro. 

\newpage
\appendix

\section{Validation on Representative Benchmarks}
\label{sec:benchmarksVAL}

In this section, we evaluate the performance of the proposed procedure analyzing four representative benchmark points. In particular, the SLQ with $m(LQ_3^{S}) = 1365$ GeV and the VLQ with $m(LQ_3^{V}) = 1850$ GeV represent completely new mass values that were not included from both the classifier training (Model A and Model B) and the statistical calibration fits ($\mu_{S,V}$ and $\sigma_{S,V}$).

The results are summarized in Table~\ref{tab:benchmark_performance}. For each benchmark point, we report three quantities derived exclusively from the output of Model~A: the mean observed discovery significance ($\langle \mathcal{Z}_{\text{obs}}\rangle$) obtained from the BL discovery test, the mean expected signal yield ($\langle S_{\text{exp}}\rangle$) extracted via a BL fit, and the corresponding inferred mass ($\langle m_{\text{inf}} \rangle$) derived from $\langle S_{\text{exp}} \rangle$. They are computed for 1000 realizations, and therefore we present their mean and one standard deviation; however, we emphasize that a single value for each observable would be obtained from real HL-LHC data. 

Furthermore, we quantify the spin identification performance of Model~B by computing the fraction of realizations classified into four mutually exclusive categories based on the dual threshold criteria defined in Sec.~\ref{modelB} ($\mathcal{Z}_{S,V} < 2$ for compatibility and $\mathcal{Z}_{S,V} > 5$ for incompatibility): correctly identified signal as scalar (SLQ id), correctly identified signal as vector (VLQ id), compatible with both hypotheses (Both), and incompatible with both hypotheses (Neither).

As shown in Table~\ref{tab:benchmark_performance}, the algorithm achieves robust identification power across all benchmark points. Crucially, the performance of the new mass points ($1365~\text{GeV}$ and $1850~\text{GeV}$) shows consistency with the interpolated expectations, confirming that the framework generalizes to intermediate masses without loss of sensitivity or classification power.

\begin{table}[ht]
\centering

\begin{tabular}{lccc cccc}
\toprule
\multirow{2}{*}{Benchmark} & \multirow{2}{*}{$\langle \mathcal{Z}_{\text{obs}} \rangle$} & \multirow{2}{*}{$\langle S_{\text{exp}} \rangle$} & \multirow{2}{*}{$\langle m_{\text{inf}} \rangle$ [GeV]} & \multicolumn{4}{c}{Identification Fractions} \\
\cmidrule(lr){5-8}
& & & & SLQ id & VLQ id & Both & Neither \\
\midrule
$\mathrm{SLQ}\ (1365~\text{GeV})$ & 17.2 $\pm$ 1.1 & 287 $\pm$ 21 & 1371 $\pm$ 14 & 0.928 & 0.0 & 0.0 & 0.072 \\
$\mathrm{SLQ}\ (1500~\text{GeV})$   & 9.6 $\pm$ 1.0 & 141 $\pm$ 17 & 1499 $\pm$ 27 & 0.907 & 0.0 & 0.0 & 0.093 \\
\midrule
$\mathrm{VLQ}\ (1850~\text{GeV})$ & 13.5 $\pm$ 1.1 & 216 $\pm$ 19 & 1823 $\pm$ 19 & 0.0 & 0.943 & 0.0 & 0.057 \\
$\mathrm{VLQ}\ (2000~\text{GeV})$   & 5.8 $\pm$ 1.0 & 101 $\pm$ 16 & 1960 $\pm$ 30 & 0.0 & 0.711 & 0.002 & 0.287 \\
\bottomrule

\end{tabular}
\caption{Performance evaluation over 1000 pseudo-experiments for four representative benchmark points. Mass values of $1365~\text{GeV}$ (SLQ) and $1850~\text{GeV}$ (VLQ) correspond to new benchmark points, not used during training or interpolation calibration.}
\label{tab:benchmark_performance}
\end{table}

\section{Exclusion limits}
\label{app:ML}

For completeness, in this appendix we show in Fig.~\ref{LQ-exc} the current expected exclusion limits from the ATLAS Collaboration~\cite{ATLAS:2020dsf,ATLAS:2021jyv,ATLAS:2023uox}, obtained from searches optimized for different final states (see the legend of the figure). We restrict the comparison to ATLAS searches, as these currently provide stronger sensitivity to the third-generation leptoquark scenario considered here than the corresponding CMS searches. We additionally show the exclusion limits obtained in our analysis using the full output distribution of Model~A for the SM--LQ discrimination, together with the BL approach and the corresponding exclusion test. Results are shown for full Run~2 (139 fb$^{-1}$), full Run~3 (300 fb$^{-1}$), and the full HL-LHC dataset (3000 fb$^{-1}$). These results complement those presented in Fig.~\ref{fig:discovery_exclusion_limits}, providing the corresponding exclusion reach for all center-of-mass energies and integrated luminosities considered in our analysis.
As expected, the exclusion limits obtained from searches targeting different final states provide complementary coverage of the parameter space. Nevertheless, most of the region excluded by ATLAS can already be accounted for by the analysis of Ref.~\cite{ATLAS:2021jyv}, which targets the same final state considered in this work. This provides a direct and consistent reference for assessing the exclusion sensitivity achieved by our ML-based approach.

\begin{figure}
  \centering
  \includegraphics[width=0.85\textwidth]{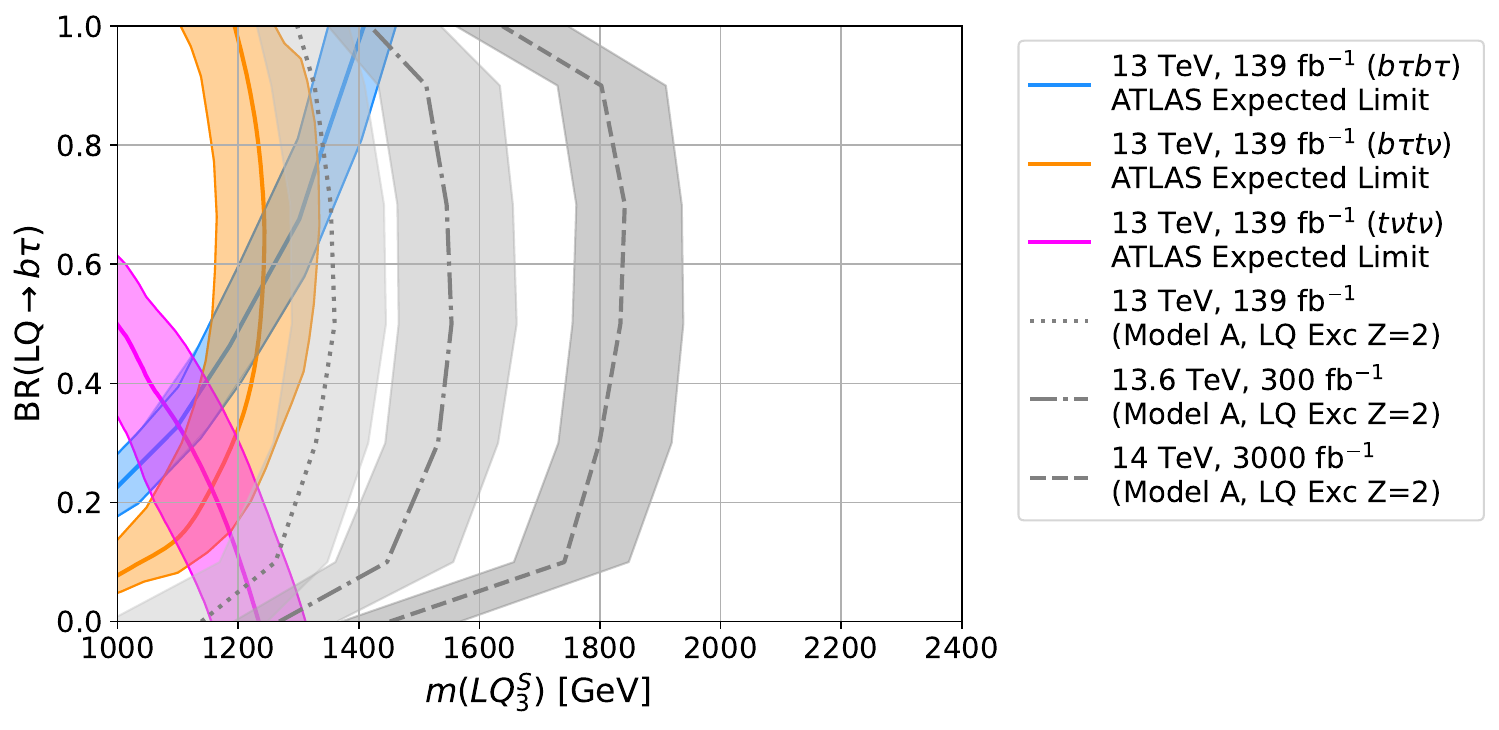}\\
  \includegraphics[width=0.85\textwidth]{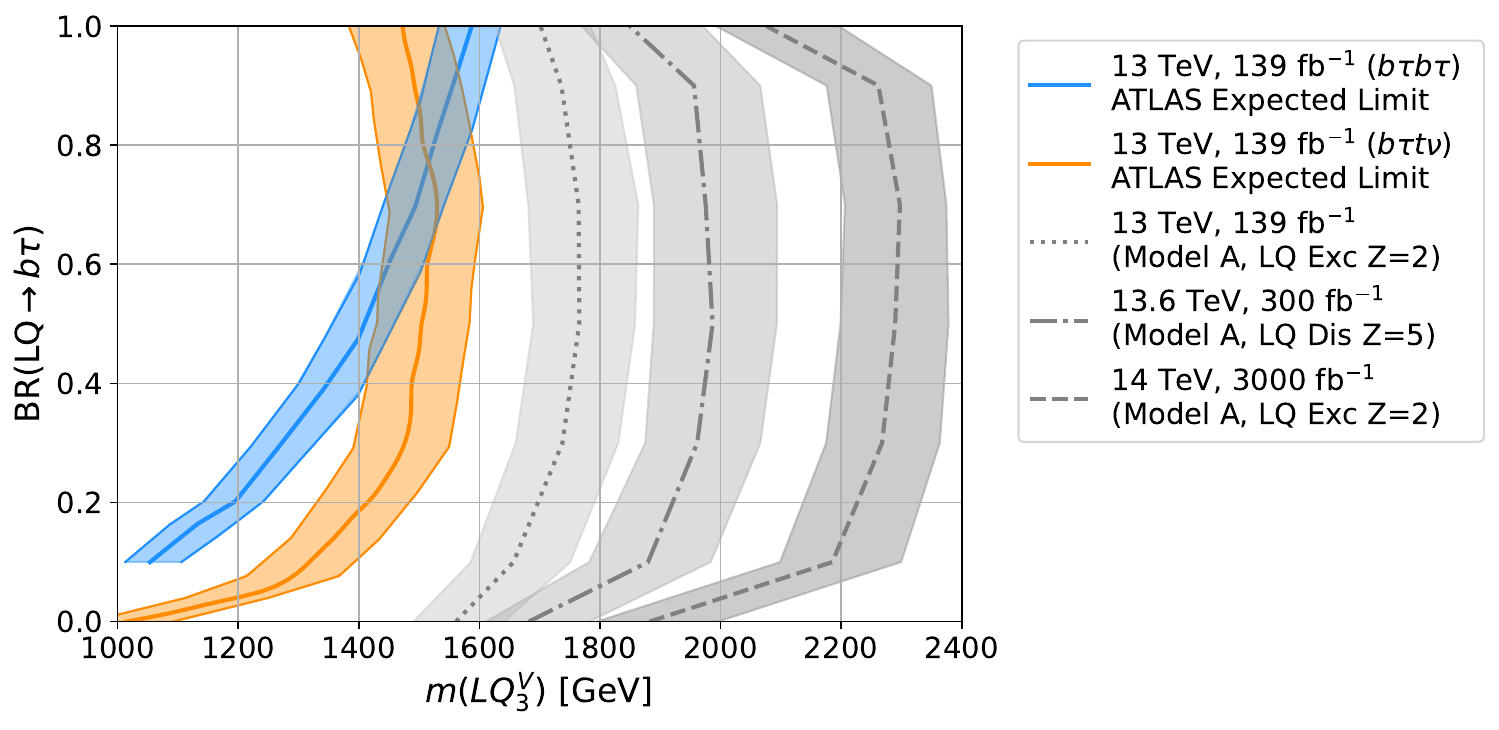}
\caption{
Same as Fig.~\ref{fig:discovery_exclusion_limits}, but showing only current and projected exclusion limits. For current ATLAS constraints~\cite{ATLAS:2020dsf, ATLAS:2021jyv,ATLAS:2023uox}, the legend indicates the final state that the search was optimized for. There is a $t\nu t\nu$ constraint as a function of $\beta$ for the SLQ case but not for the VLQ. For the limits obtained with the proposed method, optimized for $b \tau t \nu$, we show the reach for full Run 2 (139 fb$^{-1}$), full Run 3 (300 fb$^{-1}$) and for full HL-LHC (3000 fb$^{-1}$) data.
}
\label{LQ-exc}
\end{figure}

\newpage

\bibliographystyle{JHEP}
\bibliography{AdlRPSSSarXivv1.bbl}

\end{document}